\documentclass[%
 reprint,
superscriptaddress,
 amsmath,amssymb,
 aps,
]{revtex4-2}

\usepackage{graphicx}
\usepackage{dcolumn}
\usepackage{bm}
\usepackage{amsmath}
\usepackage{subcaption} 
\usepackage[justification=raggedright,singlelinecheck=false]{caption}

\begin{document}

\preprint{APS/123-QED}

\title{Uncertainty-Aware Sparse Identification of Dynamical Systems \\
via Bayesian Model Averaging}

\author{Shuhei Kashiwamura}
\affiliation{Department of Physics, Graduate School of Science, The University of Tokyo, 7-3-1 Hongo, Bunkyo-ku, Tokyo 113-0033, Japan}
\affiliation{Physics \& Informatics Laboratories, NTT Research, Inc., Sunnyvale, California 94085, USA}

\author{Yusuke Kato}%
\affiliation{Department of Complexity Science and Engineering, Graduate School of Frontier Sciences, The University of Tokyo, 5-1-5 Kashiwanoha, Kashiwa, Chiba 277-8561, Japan}
\affiliation{Department of Computational Medicine and Bioinformatics, University of Michigan, MI 48109, USA}

 \author{Hiroshi Kori}%
\affiliation{Department of Complexity Science and Engineering, Graduate School of Frontier Sciences, The University of Tokyo, 5-1-5 Kashiwanoha, Kashiwa, Chiba 277-8561, Japan}

\author{Masato Okada}%
\affiliation{Department of Complexity Science and Engineering, Graduate School of Frontier Sciences, The University of Tokyo, 5-1-5 Kashiwanoha, Kashiwa, Chiba 277-8561, Japan}
\affiliation{Department of Physics, Graduate School of Science, The University of Tokyo, 7-3-1 Hongo, Bunkyo-ku, Tokyo 113-0033, Japan}




\date{\today}

\begin{abstract}
In many problems of data-driven modeling for dynamical systems, the governing equations are not known a priori and must be selected phenomenologically from a large set of candidate interactions and basis functions. In such situations, point estimates alone can be misleading, because multiple model components may explain the observed data comparably well, especially when the data are limited or the dynamics exhibit poor identifiability. Quantifying the uncertainty associated with model selection is therefore essential for constructing reliable dynamical models from data.
In this work, we develop a Bayesian sparse identification framework for dynamical systems with coupled components, aimed at inferring both interaction structure and functional form together with principled uncertainty quantification. The proposed method combines sparse modeling with Bayesian model averaging, yielding posterior inclusion probabilities that quantify the credibility of each candidate interaction and basis component.
Through numerical experiments on oscillator networks, we show that the framework accurately recovers sparse interaction structures with quantified uncertainty, including higher-order harmonic components, phase-lag effects, and multi-body interactions. We also demonstrate that, even in a phenomenological setting where the true governing equations are not contained in the assumed model class, the method can identify effective functional components with quantified uncertainty. These results highlight the importance of Bayesian uncertainty quantification in data-driven discovery of dynamical models.

\end{abstract}

\maketitle


\section{Introduction}
Dynamical phenomena in many areas of science are described using mathematical models whose governing equations must be inferred from observations \cite{strogatz2024nonlinear}. In practice, constructing such models amounts to determining which terms should appear in the equations among many possible candidates \cite{brunton2016discovering}. This identification involves two complementary aspects. One concerns the interaction structure, namely which variables influence each other, as in network reconstruction problems \cite{timme2014revealing}. The other concerns the functional form of the dependence, namely how the influence depends on the system state, as studied in equation discovery and system identification approaches \cite{ljung2010perspectives, schmidt2009distilling}. Although these problems are often treated separately, both can be understood as selecting relevant components from a candidate representation of the dynamics.

Dynamical systems composed of interacting components are ubiquitous in both natural and engineered contexts. They appear in ecological communities, neural populations, coupled mechanical and electrical circuits, and oscillator networks underlying circadian rhythms \cite{bairey2016high,dayan2005theoretical,izhikevich2007dynamical,strogatz1987human}, as well as in chemical reaction and cardiac systems \cite{horn1972general,feinberg1987chemical,martynenko2022mathematical}. In many of these cases both the interaction structure and the functional form of the interactions are unknown and must be inferred from data, such as reaction rate laws, regulatory response functions, or nonlinear coupling terms. Identifying these components enables understanding and control of the dynamics, such as revealing synchronization mechanisms in biological systems, designing control strategies in engineered networks, and characterizing influence patterns in ecological or neural systems \cite{winfree1967biological,leleu2021scaling,banerjee2023network,kori2008synchronization,kiss2007engineering}.

A wide variety of data driven methods have been proposed to infer dynamical models from observed time series \cite{
stankovski2017coupling,
rosenblum2023inferring,
rosenblum2001detecting,
tokuda2007inferring,
kralemann2007uncovering,
kralemann2011reconstructing,
stankovski2012inference,
stankovski2015coupling,
stankovski2019coupling,
ota2014direct,
ota2020interaction,
tokuda2019practical,
tabar2024revealing,
matsuki2025network,
timme2014revealing}.
Among these approaches, many rely on linear estimation procedures. When the dynamics can be expressed as a linear combination of candidate basis functions, the task reduces to estimating coefficients within a linear regression framework, which simultaneously determines which variables interact and which functional components are required.
This formulation is broadly applicable because nonlinear vector fields can often be approximated or expanded using suitable basis functions such as polynomial, trigonometric, or radial functions. Typical examples include oscillator networks such as the Kuramoto model \cite{acebron2005kuramoto,kuramoto2003chemical}, in which the interaction of oscillators is described by trigonometric functions, coupled biological or chemical systems governed by polynomial equations, and chemical reaction networks described by parametric rate laws \cite{horn1972general}.

Within this class of linear estimation approaches, sparse estimation plays a central role. 
In many dynamical systems, only a small subset of possible interactions is active \cite{himeoka2025perturbation}, and the appropriate functional form of these interactions is often unknown in advance. 
Thus, sparsity is important not only for selecting the interacting elements but also for identifying, from a potentially large dictionary, the subset of basis functions that actually governs the dynamics \cite{brunton2016discovering,fasel2022ensemble,hirsh2022sparsifying,bertsimas2023learning}.
Methods such as the least absolute shrinkage and selection operator (LASSO) \cite{mangan2017inferring} and adaptive LASSO \cite{su2025pairwise} exploit this structure by jointly estimating coefficients and selecting relevant interactions or basis functions, and have been widely used to infer network structures from time-series data.
However, conventional sparse regression approaches provide only point estimates and do not quantify the uncertainty associated with inferred connections or selected basis functions, which can lead to overconfidence or misinterpretation when data are limited or the system is high-dimensional.

Bayesian inference offers a principled framework for uncertainty quantification \cite{gelman1995bayesian}. It yields posterior distributions over model parameters conditioned on observed data, providing probabilistic measures of confidence in the inferred results \cite{kato2025bayesian}. Bayesian methods for sparse estimation or variable selection have played an important role in fields such as economics \cite{wang2017model}, psychology \cite{bainter2023comparing}, health science \cite{garcia2021bayesian}, and materials science \cite{obinata2022data,obinata2025confidence}. By examining posterior distributions, one can assess how credible each candidate variable is to be nonzero, thereby obtaining a natural uncertainty measure for selecting relevant model components.

In this work, we adopt a Bayesian approach for sparse identification of dynamical models whose vector fields are expressed as linear combinations of candidate basis functions. We introduce binary indicator variables that represent the presence or absence of each candidate term in the governing equations. By marginalizing over the model parameters, we obtain posterior inclusion probabilities for these indicators, which quantify how likely each component contributes to the dynamics. This corresponds to Bayesian model averaging (BMA) \cite{raftery1997bayesian,wasserman2000bayesian,fragoso2018bayesian,bishop2006pattern} and provides an interpretable measure of uncertainty over competing model representations.

We demonstrate the effectiveness of this framework using oscillator‑based systems.
For general networks of phase oscillators characterized by trigonometric coupling functions—of which the Kuramoto model is a representative example—the method successfully identifies pairwise and higher‑order interactions as well as harmonic components, allowing interpretable reconstruction of sparse coupling structures.

Beyond such settings where the model class is known, the same framework can be used for phenomenological modeling when the governing equations are not explicitly specified. By approximating the dynamics of a mechanical metronome system with a reduced phase representation, the method determines which functional components are required to explain the observed motion and provides uncertainty aware data driven phase reduction. These results illustrate that the proposed Bayesian approach is applicable both to identifying interactions within a given model class and to constructing effective dynamical models from observations.

\begin{figure}[t] 
  \centering
  \includegraphics[width=\columnwidth]{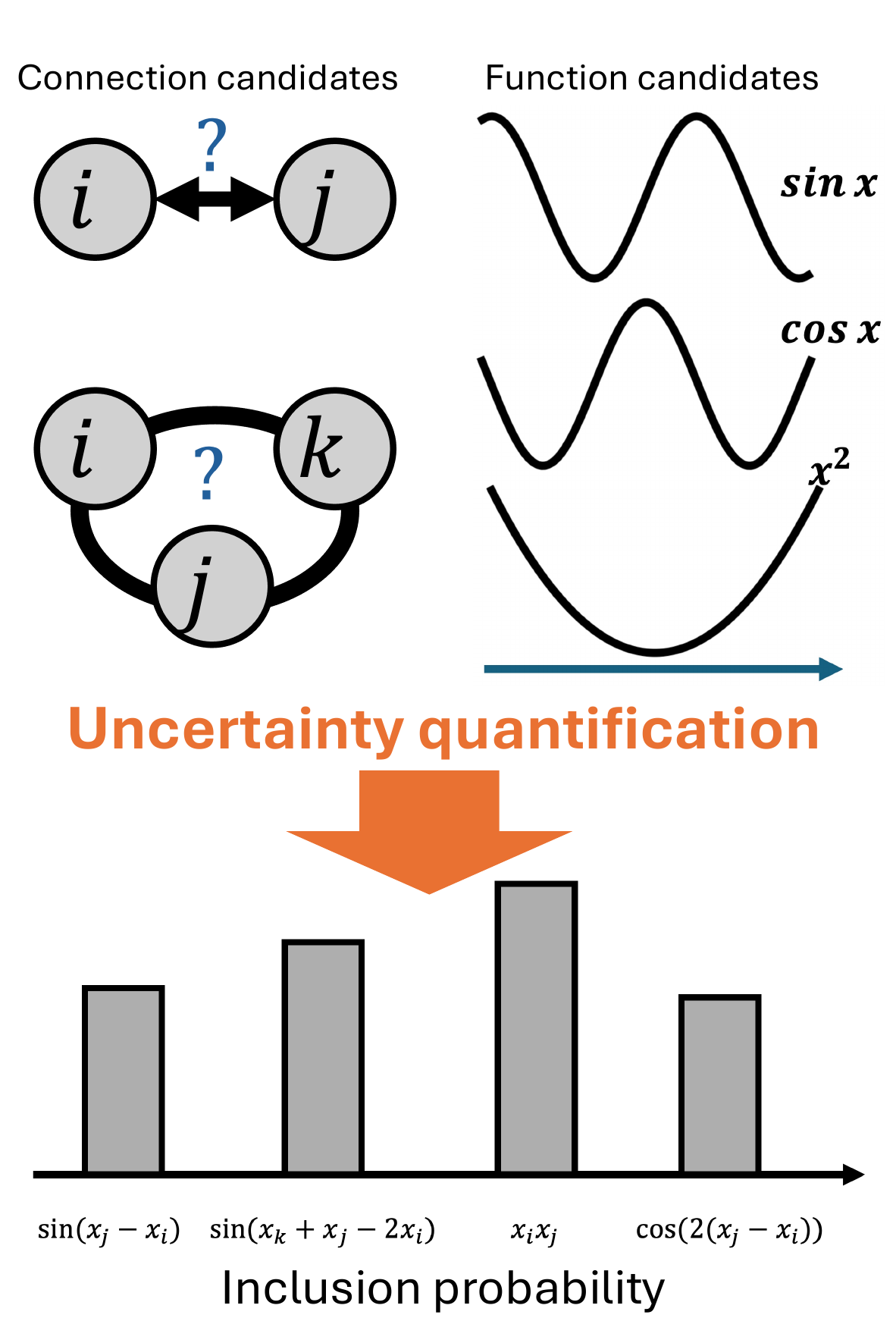}
  \caption{Illustration of our approach.
We infer the interaction structure of a dynamical system, as well as the functional forms, from a set of candidates.
Within a Bayesian framework, we quantify uncertainty by computing inclusion probabilities via Bayesian model averaging.}
  \label{fig:diagram}
\end{figure}

\section{Class of Dynamical Systems Considered}\label{sec2}
The scope of this study covers a broad class of dynamical systems whose state variables are represented by
\begin{equation}
  \bm{X} = (X_1, X_2, \dots, X_N)^\top \in \mathbb{R}^N,
\end{equation}
and whose time evolution can be expressed as a linear combination of basis functions:
\begin{equation}
  \frac{dX_i}{dt}
  = \sum_{e\in E_i} a_e F_e(\bm{X}_e).
  \label{eq:general_dynamics}
\end{equation}
Here \( e \) does not represent a simple index but a generalized edge representing connections between multiple components and \(\bm{X}_e\) represents the collection of state variables involved in that interaction.
Then, this notation includes wide range of interactions such as directed two body connection $e = (i\rightarrow j)$, undirected connection $e=(i,j)$, three-body connection $e=(i,j,k)$ and so on.
And $E_i$ denotes all of edges that affects to elements $X_i$.
The function \(F_e(\bm{X}_e)\) defines the functional form of the interaction,
while the coefficient \(a_e\) quantifies its strength.
This formulation provides a unified representation encompassing a wide range of nonlinear dynamical systems,
since nonlinearities can be captured by selecting appropriate basis functions \(F_e\).

Prominent examples of systems within this class include oscillator networks,
where trigonometric basis functions of phase differences such as
\begin{equation}
\frac{dX_i}{dt}=\omega_i+\sum_j \sum_l\sin(l(X_j - X_i) + \alpha_{ij}),
\end{equation}
where $\alpha_{ij}$ denotes the phase lag, $\omega_i$ the natural frequency, and $l\in \mathbb Z$ the harmonic order. This interaction appear in the Kuramoto or Sakaguchi--Kuramoto models \cite{kuramoto2003chemical}.
These can be further extended to many body interactions, for example,
\(\sin(X_j + X_k - 2X_i)\) or \(\sin(2X_j - X_k - X_i)\),
where each edge \(e\) connects three oscillators \((i, j, k)\).
Other representative instances are the Lotka--Volterra systems for population dynamics \cite{wangersky1978lotka},
\begin{equation}
  \frac{dX_i}{dt} = X_i\left(r_i + \sum_j a_{ij} X_j\right),
\end{equation}
and gene regulatory networks \cite{karlebach2008modelling},
\begin{equation}
  \frac{dX_i}{dt} = \sum_j a_{ij} g_{ij}(X_j),
\end{equation}
where \(g(X_j)\) is a nonlinear activation function such as a Hill function describing regulation or inhibition.
All of these examples can be expressed in the general form of Eq.~\eqref{eq:general_dynamics} by choosing suitable basis functions \(F_e\).

This general formulation encompasses oscillator, ecological, neural, and biochemical systems.
In practice, the governing functional forms of the dynamics are often unknown prior to analysis.
Instead, one specifies a dictionary of candidate basis functions $F_e$ and infers which of them actually contribute to the observed dynamics.
Therefore, the problem is not only to determine the interaction structure represented by the edges $e$, but also to identify the subset of functional components required to describe the system.

\section{Methods}\label{sec3}
\subsection{Data}

We consider a continuous-time dynamical system of $N$ elements with state trajectories
$X_i(t)$, $i=1,\dots,N$.
These trajectories are observed at uniform time intervals $\Delta t$, yielding the sampled sequence
\begin{align}
\bm X_i 
&= \left( X_{i,0}, \, X_{i,1}, \, \ldots, \, X_{i,M} \right)
    \in \mathbb{R}^{M+1}, \\
X_{i,m} 
&= X_i(m\Delta t),
\qquad m = 0,1,\ldots,M.
\end{align}

From the sampled trajectory, we define the target signal as the discrete difference of states:
\begin{align}
Y_{i,m}
&:= X_{i,m+1} - X_{i,m},
\qquad m = 0,1,\ldots,M-1.
\end{align}

Consequently, for each element $i$, we obtain the data vector
\begin{align}
\bm Y_i
&=
\bigl( Y_{i,0}, \, Y_{i,1}, \, \ldots, \, Y_{i,M-1} \bigr)^\top
\in \mathbb{R}^{M}.
\end{align}

Thus, the complete dataset is given by
\begin{align}
D := \{ \bm Y_i \}_{i=1}^N,
\end{align}
where each $\bm Y_i$ represents the observed sequence of state differences for element $i$.

\subsection{Sparse modeling for network structure}
We model the difference of state $\Delta X_{i,m}$ by linear combination of basis functions
$\{g_{i,m}^{(\gamma)}(\bm X_i^{(\gamma)})\}_{\gamma=1}^{\Gamma}$, where
$\bm X_i^{(\gamma)} \subset \bm X_i$ represents the subset of elements which has connection
through interaction $g_{i,m}^{(\gamma)}(\bm X_i^{(\gamma)})$.
Note that the notation $\bm X_i^{(\gamma)}$ does not refer to the state of element $i$ itself,
but to the collection of state variables of elements that influence $i$ through interaction $\gamma$.
We omit the $\bm X_i^{(\gamma)}$ and denote it as $g_{i,m}^{(\gamma)}$.
This set of basis functions should include all of possible candidates we consider.

We introduce the binary variables $c_i^{(\gamma)} \in \{0,1\}$, which is called indicator, determining whether basis function $g_{i,m}^{(\gamma)}$ is included or not. 
Indicator variables $c_i^{(\gamma)} $ also describe the structure of interactions because it determines the subset $\bm X_i^{(\gamma)}$ is connected or not.
Then, we have the following formulation,
\begin{align}
\Delta X_{im}
  &= X_{i,m+1} - X_{i,m} \notag \\
  &= \Delta t \sum_{\gamma=1}^\Gamma 
      c_i^{(\gamma)} \Theta_i^{(\gamma)} g_{i,m}^{(\gamma)}  \\
  &= \Delta t\, (\bm c_i \odot \bm \Theta_i)^\top \bm g_{i,m},
  \label{general_formulation}
\end{align}
where each vector is defined as 
$\bm c_i = \{c_i^{(\gamma)}\}_{\gamma=1}^{\Gamma}$, 
$\bm \Theta_i = \{\Theta_i^{(\gamma)}\}_{\gamma=1}^{\Gamma}$, 
and 
$\bm g_{i,m}=\{g_{i,m}^{(\gamma)}\}_{\gamma=1}^{\Gamma}$, 
with the assumption that the discrete update
$\Delta X_{i,m}$ follows the forward Euler approximation.
Here $\odot$ denotes the Hadamard product.

Under the assumption that the continuous-time dynamics are perturbed by
additive Gaussian white noise and that the observation process also
introduces independent Gaussian noise, the discrete-time differences
$\bm Y_i$ are modeled with an effective noise variance
$\sigma_i^2 = (\sigma_i^{(d)})^2 \Delta t + (\sigma_i^{(o)})^2$, where $(\sigma_i^{(d)})^2$ and $(\sigma_i^{(o)})^2$ denote the variance of dynamical white noise and observational noise, respectively.
We define the likelihood for the data of each element as follows:
\begin{align}
&P(\bm Y_i \mid \bm \Theta_i, \bm c_i, \sigma_i) = \notag \\
&\left(\frac{1}{2\pi\sigma_i^2}\right)^{M/2}
\exp\!\left(
  -\frac{1}{2\sigma_i^2}
  \left\|\bm Y_i - \Delta t\, \bm G_i (\bm c_i \odot \bm \Theta_i)\right\|^2
\right),
\end{align}
where $\bm G_i\in\mathbb R^{M\times\Gamma}$ is the design matrix with
$G_{i,m,\gamma}=g_{i,m}^{(\gamma)}$.

Under this formulation, identifying the governing dynamics reduces to a sparse linear regression problem over candidate model components. We employ a Bayesian approach to quantify the credibility of the inferred sparsity pattern represented by the indicator variables $\bm c_i$. 

\subsection{Bayesian model}

In our framework, the indicator variables $\bm{c}_i$, the parameter vectors $\bm{\Theta}_i$, and the hyperparameters governing their prior distributions are treated as random variables, and our objective is to infer the posterior distributions of $\bm{c}_i$ and $\bm{\Theta}_i$ conditioned on the observed data $\bm{Y}_i$.
We assume a hierarchical Bayesian framework in which the indicator $\bm{c}_i$ is first drawn from a prior distribution $P(\bm{c}_i)$, the parameter vector $\bm{\Theta}_i$ is then sampled from a conditional prior $P(\bm{\Theta}_i \mid \bm{c}_i)$, and the observed data $\bm{Y}_i$ are generated from the likelihood $P(\bm{Y}_i \mid \bm{\Theta}_i, \bm{c}_i)$.
In this formulation, $\bm{c}_i$ encodes the presence or absence of candidate terms determining the model structure, while $\bm{\Theta}_i$ represents the corresponding coefficients.

We define the prior distributions as
\begin{align}
P(\bm c_i)
  &= \prod_{\gamma=1}^{\Gamma}
     p^{c_i^{(\gamma)}}\left(1-p\right)^{1-c_i^{(\gamma)}}, \label{eq:prior_c}
\end{align}
\begin{align}
P(\bm{\Theta}_i \mid \bm{c}_i, \bm{\tau}_i)
= \prod_{\gamma=1}^{\Gamma}
\Biggl[
&\, c_i^{(\gamma)} \,
\frac{1}{\sqrt{2\pi \tau_{i\gamma}^2}}
\exp\!\left(-\frac{\Theta_{i\gamma}^2}{2\tau_{i\gamma}^2}\right) \nonumber \\
&+ (1 - c_i^{(\gamma)}) \, \delta(\Theta_{i\gamma})
\Biggr].\label{eq:prior_theta}
\end{align}

Equation~\eqref{eq:prior_c} is a Bernoulli prior with inclusion probability $p$, while Eq.~\eqref{eq:prior_theta} represents a spike-and-slab prior:
when $c_i^{(\gamma)}=1$, the coefficient $\Theta_{i\gamma}$ follows a zero-mean Gaussian with variance $\tau_{i\gamma}^2$;
otherwise, it is fixed to zero via the delta function.

Given our goal of inferring the indicator vector $\bm c_i$,
we base our inference on the posterior distribution of $(\bm c_i, \sigma_i, \bm \tau_i)$
conditioned on the observed data $\bm Y_i$.
By Bayes' theorem, this posterior is given by
\begin{align}
P(\bm c_i, \sigma_i, \bm \tau_i \mid \bm Y_i)
 = \frac{P(\bm Y_i \mid \bm c_i, \sigma_i, \bm \tau_i)\,
         P(\bm c_i, \sigma_i, \bm \tau_i)}
        {P(\bm Y_i)}, \label{eq:posterior}
\end{align}
where the normalization constant is
\begin{align}
P(\bm Y_i)
 = \sum_{\bm c_i}\!
   \int d\sigma_i\, d\bm \tau_i\,
   P(\bm Y_i \mid \bm c_i, \sigma_i, \bm \tau_i)
   P(\bm c_i, \sigma_i, \bm \tau_i). \label{eq:evidence}
\end{align}

The joint prior $P(\bm c_i, \sigma_i, \bm \tau_i)$ reflects our modeling assumptions
about the sparsity pattern, noise level, and coefficient scales, and is chosen to factorize as
\begin{align}
P(\bm c_i, \sigma_i, \bm \tau_i)
 = P(\bm c_i)\, P(\sigma_i)\, P(\bm \tau_i), \label{eq:prior_factorize}
\end{align}
where $P(\bm c_i)$ is the Bernoulli prior defined above, and
$P(\sigma_i)$ and $P(\bm \tau_i)$ are taken to be uniform over specified ranges.

The remaining ingredient in \eqref{eq:posterior} is the marginal likelihood
$P(\bm Y_i \mid \bm c_i, \sigma_i, \bm \tau_i)$.
This quantity can be derived in closed form by analytically integrating out the coefficients
$\bm \Theta_i$, as 
\begin{align}
P(\bm Y_i|\bm c_i, \sigma_i, \bm \tau_i)
 = \int d\bm{\Theta}_i\,
    P(\bm Y_i \mid \bm{\Theta}_i, \bm c_i, \sigma_i)\,&
    P(\bm{\Theta}_i \mid \bm c_i, \bm \tau_i) \notag\\
 = \int d\bm{\Theta}_i^c\,
    \mathcal{N}\!\left(\bm Y_i \mid \Delta t\, \bm G_i^c \bm \Theta_i^c,\, \sigma_i^2 I\right)&
    \mathcal{N}\!\left(\bm \Theta_i^c \mid \bm 0,\, \Lambda_i^c \right) \notag\\
 = \mathcal{N}\!\left(
      \bm Y_i \mid \bm 0,\,
      \sigma_i^2 I_M
      + \Delta t^2\, \bm G_i^c \Lambda_i^c {\bm G_i^c}^\top
    \right)&. \label{eq:marginal_likelihood}
\end{align}
Here, $\mathcal{N}(\cdot \mid \bm\mu, \Sigma)$ denotes a multivariate normal distribution with mean $\bm\mu$ and covariance $\Sigma$.
$\bm G_i^c \in \mathbb{R}^{M\times|\bm c_i|_0}$ contains only the columns of the design matrix $\bm G_i$ corresponding to active components ($c_i^{(\gamma)}=1$);
$\bm \Theta_i^c\in\mathbb{R}^{|\bm c_i|_0}$ is the corresponding parameter vector;
and $\Lambda_i^c = \mathrm{diag}(\{\tau_{i\gamma}^2: c_i^{(\gamma)}=1\})$ is a diagonal matrix of prior variances.

Given this expression, the posterior distribution 
$P(\bm c_i, \sigma_i, \bm \tau_i \mid \bm Y_i)$ 
can be evaluated up to a normalization constant for each indicator $\bm c_i$.
The posterior inclusion probability of each  term,
which quantifies the credibility that $c_i^{(\gamma)} = 1$,
is obtained as
\begin{align}
\Pr(c_i^{(\gamma)} = 1 \mid \bm Y_i)
  = \sum_{\bm c_i} c_i^{(\gamma)}\, P(\bm c_i \mid \bm Y_i), 
\label{eq:bma}
\end{align}
where $P(\bm c_i \mid \bm Y_i)$ is the marginal posterior distribution 
after integrating out $\sigma_i$ and $\bm \tau_i$.
This quantity corresponds to Bayesian model averaging (BMA) 
and provides a natural measure of uncertainty in the inferred model structure.

To obtain a point estimate of the model structure,
one can apply a threshold (for example $0.5$) to the posterior inclusion probabilities
$\Pr(c_i^{(\gamma)} = 1 \mid \bm Y_i)$
and determine the binary estimator $\hat{\bm c}_i$ accordingly.

Once the structure $\hat{\bm c}_i$ is fixed, 
the conditional posterior distribution of the active coefficients
$\bm \Theta_i^c$ is multivariate Gaussian,
\[
P(\bm \Theta_i^c \mid \bm Y_i, \hat{\sigma}_i, \Lambda_i^c)
= \mathcal{N}\!\left(
    \bm \Theta_i^c \mid 
    \hat{\bm \Theta}_i^{c},\, 
    \Sigma_i
  \right),
\]
with posterior mean and covariance given by
\begin{align}
\hat{\bm \Theta}_i^{c}
&= \frac{\Delta t}{\hat{\sigma}_i^2}\,
    \Sigma_i\, {G_i^c}^\top \bm Y_i,  \label{eq:coeff_esti}\\
\Sigma_i
&= \left(
    \frac{\Delta t^2}{\hat{\sigma}_i^2}\,{G_i^c}^\top G_i^c
    + (\Lambda_i^c)^{-1}
   \right)^{-1}.
\end{align}
These characterize the most probable coupling strengths and their 
associated uncertainty.

Additionally, the number $\Gamma$ of basis functions can be treated as a random variable, leading to the extended posterior
\begin{equation}
P(\bm c_i, \sigma_i, \bm \tau_i, \Gamma \mid \bm Y_i).
\end{equation}
This formulation not only infers the underlying interaction structure but also performs sparse selection of the most plausible basis functions from a candidate dictionary.
Assuming that the time--series data of different elements $\{X_i\}$ are conditionally independent given their parameters, 
the joint posterior over all elements factorizes as
\begin{align}
P\bigl(\{\bm c_i, \sigma_i, \bm \tau_i\}_{i=1}^N, \Gamma \,\big|\, \{\bm Y_i\}_{i=1}^N \bigr)
&=
\prod_{i=1}^N 
P(\bm c_i, \sigma_i, \bm \tau_i, \Gamma \mid \bm Y_i).
\end{align}
This assumption is reasonable in the present formulation, where the inference is performed in a node-wise regression manner.
However, for modeling scenarios that involve shared parameters across nodes (e.g., common intrinsic frequencies) or explicitly coupled interaction structures, a formulation that accounts for dependencies among nodes would be required.

\section{Parallel Tempering}\label{sec4}
Direct evaluation of the posterior distribution in Eq.~(\ref{eq:posterior}) 
is challenging because its normalization constant in Eq.~(\ref{eq:evidence}) 
requires a high-dimensional integration over the hyperparameters 
and a combinatorial summation over the indicator variables.
Furthermore, computing expectations, such as that of Eq.~(\ref{eq:bma}) in closed form 
is analytically intractable.
To obtain these quantities numerically, we resort to sampling from the 
posterior distribution.

However, the discrete indicators $\bm c_i$ induce a highly rugged and combinatorial posterior landscape. 
As a consequence, standard Markov chain Monte Carlo (MCMC) methods based on 
local proposals tend to mix slowly and can easily become trapped in local 
regions of the parameter space.

To explore the posterior distribution more effectively, we employ 
parallel tempering \cite{hukushima1996exchange, swendsen1986replica, hansmann1997parallel, sambridge2014parallel}. 
This scheme constructs a sequence of tempered distributions with reduced 
density concentration and allows exchanges between their Markov chains, 
thereby mitigating slow mixing and alleviating trapping in local regions.

\subsection{Replica ladder and tempered posteriors}
Parallel tempering is based on introducing a family of tempered
posterior distributions parameterized by an inverse temperature
$\beta \in [0,1]$:
\[
P_{\beta}(\bm x)
  \propto 
  P(\bm Y_i \mid \bm x)^{\beta}\, P(\bm x),
\]
where $\bm x$ collectively denotes
$(\bm c_i,\sigma_i,\bm \tau_i,\Gamma)$.
When $\beta = 1$, this distribution coincides with the target posterior,
whereas $\beta = 0$ yields the prior distribution.
As $\beta$ decreases, the likelihood term becomes flatter, enabling
broader exploration of the parameter space.

To implement this idea, we construct a sequence of $R$ replicas,
each associated with a different inverse temperature
\[
0=\beta_1 < \beta_2 < \cdots < \beta_R = 1.
\]
Replica $r$ samples from the tempered distribution $P_{\beta_r}(\bm x)$:
low–inverse-temperature replicas (small $\beta_r$) explore broadly,
while the highest–inverse-temperature replica ($\beta_R = 1$)
targets the true posterior.

The temperature ladder $\{\beta_r\}$ is chosen as a geometric sequence
for replicas $r=2,\dots,R$:
\[
\beta_r = \eta^{\,r-R},
\]
with $\beta_R = 1$ fixed and $\beta_1 = 0$ manually set.
In other words, the geometric progression determines all intermediate
temperatures except for the anchor point $\beta_1=0$.
This construction is known to yield approximately constant exchange
acceptance probabilities across adjacent replicas \cite{nagata2008asymptotic}.

\subsection{Local updates and replica exchange}

Within each replica, we perform standard MCMC updates under the tempered
posterior $P_r(\bm x)$.
The indicator variables $\bm c_i$ are updated by proposing flips of the
binary components $c_i^{(\gamma)}$ and accepting or rejecting them according
to the Metropolis--Hastings rule based on $P_r$.
The continuous hyperparameters $(\sigma_i,\bm\tau_i)$ are
updated by suitable proposal distributions, again accepted or rejected
with respect to the tempered posterior.
The number of basis functions $\Gamma$ is updated by addition $\Gamma +1$ or subtraction $\Gamma-1$.
Because each replica corresponds to a different inverse temperature, the
resulting ensemble of chains explores the posterior landscape with
different levels of ruggedness. High-temperature replicas explore broadly,
while low-temperature replicas concentrate on modes of the target posterior.

For these replicas, neighboring pairs $(r,r+1)$ periodically attempt
to exchange their states $(\bm x_r,\bm x_{r+1})$.
The proposed swap is accepted with probability
\begin{align}
 \min \left[1, \frac{P_r(x_{r+1})P_{r+1}(x_r)}{P_r(x_{r})P_{r+1}(x_{r+1})}\right].
\end{align}
This exchange step preserves detailed balance with respect to the joint
distribution over all replicas and allows low-temperature chains to
escape local modes by occasionally adopting configurations generated by
high-temperature chains.
Conversely, high-temperature chains inherit information about the modes
identified by low-temperature chains, leading to an overall improvement
in global exploration.

\begin{figure*}[t]
  \centering
  \begin{subfigure}{0.32\textwidth}
    \centering
    \caption{}
    \includegraphics[width=\linewidth]{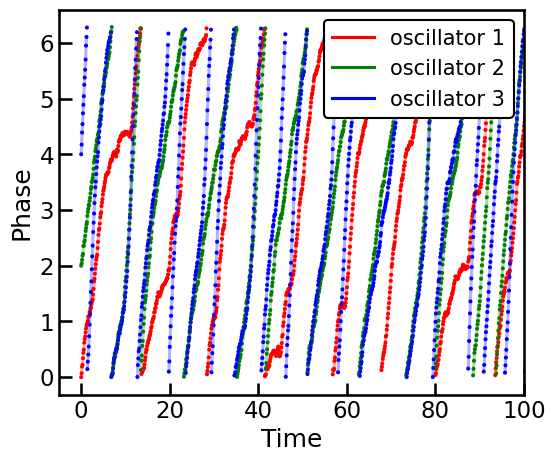}
    \label{fig:phase1}
  \end{subfigure}
  \hfill
  \begin{subfigure}{0.32\textwidth}
    \centering
    \caption{}
    \includegraphics[width=\linewidth]{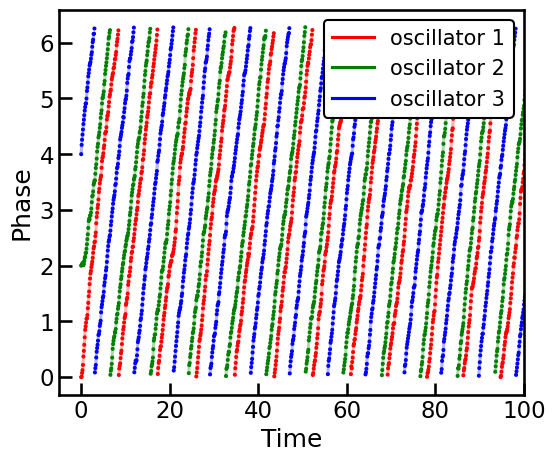}
    \label{fig:phase2}
  \end{subfigure}
  \hfill
  \begin{subfigure}{0.32\textwidth}
    \centering
    \caption{}
    \includegraphics[width=\linewidth]{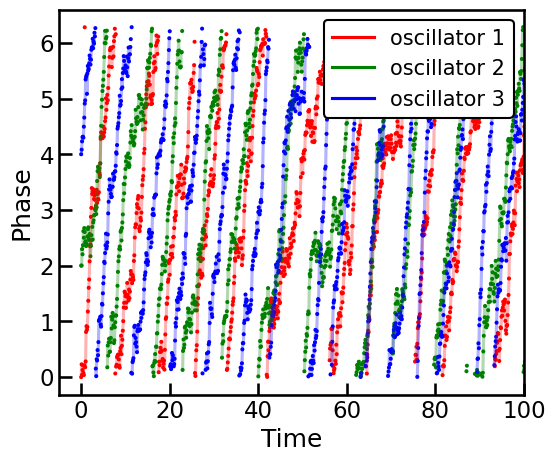}
    \label{fig:phase3}
  \end{subfigure}

  \caption{Synthetic phase time-series data $\{\bm X_i \in \mathbb R^M\}_{i=1}^N$ for $N=3$ oscillators. 
The underlying time interval is $t \in [0,200]$ with sampling interval $\Delta t = 0.1$, so that the total number of data points is $M=2000$. 
For visual clarity, only the first $1000$ points corresponding to $t \in [0,100]$ are plotted. 
The red, green, and blue dots represent the data of oscillators $i=1,2,3$, respectively, and the shaded lines connect successive points for each oscillator. 
The network structure, coupling strengths, and initial conditions are identical across panels (a)–(c). 
(a) Configuration 1: asynchronous trajectory with phase lags $\alpha_{ij}=1.0$, $\alpha_{ijk}=1.0$, and $\alpha_{ijk}'=1.0$, harmonic orders $L^{(2)}=1$ and $L^{(3)}=1$, intrinsic frequencies $(\omega_1, \omega_2, \omega_3)^{\mathsf T}
=(0.5, 1.0, 1.5)^{\mathsf T}$, and dynamical noise $\sigma_d=0.1$. 
(b) Configuration 2: phase-locked trajectory with $\alpha_{ij}=1.0$, $\alpha_{ijk}=1.0$, $\alpha_{ijk}'=0.0$, $L^{(2)}=1$, $L^{(3)}=2$,  $(\omega_1, \omega_2, \omega_3)^{\mathsf T}
=(0.4, 0.8, 1.2)^{\mathsf T}$, and $\sigma_d=0.1$. 
(c) Configuration 3: disturbed trajectory with dynamical noise $\sigma_d=0.5$ and otherwise the same parameters as in Configuration~2.
}
  \label{fig:three-phases}
\end{figure*}
\begin{figure*}[t]
  \centering
  \begin{subfigure}{0.32\textwidth}
    \centering
    \caption{}
    \includegraphics[width=\linewidth]{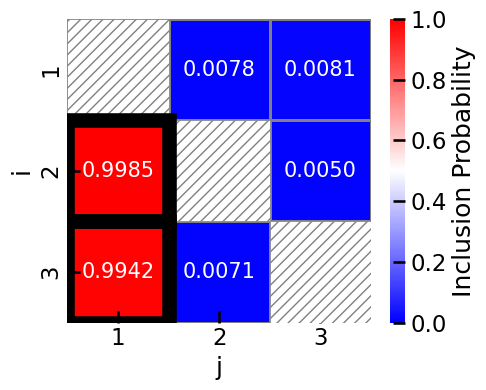}
    \label{fig:cij1}
  \end{subfigure}
  \hfill
  \begin{subfigure}{0.32\textwidth}
    \centering
    \caption{}
    \includegraphics[width=\linewidth]{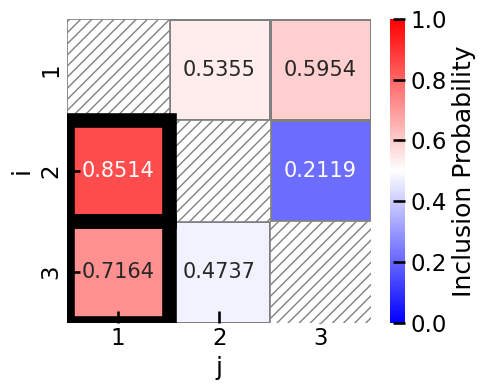}
    \label{fig:cij2}
  \end{subfigure}
  \hfill
  \begin{subfigure}{0.32\textwidth}
    \centering
    \caption{}
    \includegraphics[width=\linewidth]{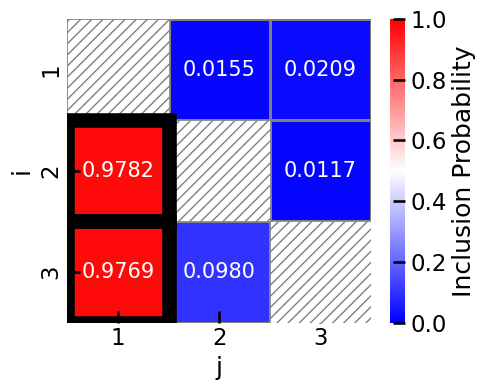}
    \label{fig:cij3}
  \end{subfigure}

  \caption{Posterior inclusion probabilities of two-body interaction indicators $c_{ij}$. 
Each panel shows the $N\times N$ matrix of inclusion probabilities for all ordered pairs $(i,j)$ with $i\neq j$, where higher values indicate stronger evidence for the presence of an interaction from oscillator $j$ to $i$. 
Self-interactions ($c_{ii}$) are not included because the model assumes no diagonal components. 
In all panels, the entries corresponding to the true nonzero interactions $\bar{c}_{21}$ and $\bar{c}_{31}$ are highlighted with thick black frames. 
(a) Results for Configuration~1 (asynchronous dynamics). 
(b) Results for Configuration~2 (phase-locked dynamics). 
(c) Results for Configuration~3 (noise-perturbed phase-locked dynamics).
}

  \label{fig:cij}
\end{figure*}

\begin{figure*}[t]
  \centering
  \begin{subfigure}{0.32\textwidth}
    \centering
    \caption{}
    \includegraphics[width=\linewidth]{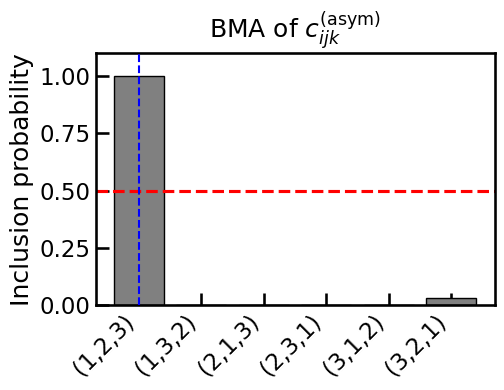}
    \label{fig:cijk_a_1}
  \end{subfigure}
  \hfill
  \begin{subfigure}{0.32\textwidth}
    \centering
    \caption{}
    \includegraphics[width=\linewidth]{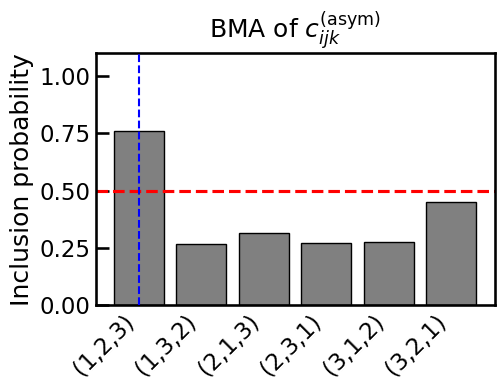}
    \label{fig:cijk_a_2}
  \end{subfigure}
  \hfill
  \begin{subfigure}{0.32\textwidth}
    \centering
    \caption{}
    \includegraphics[width=\linewidth]{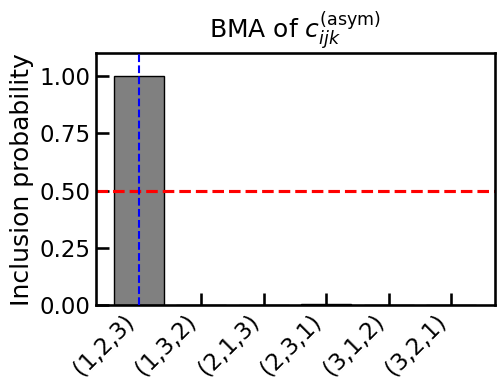}
    \label{fig:cijk_a_3}
  \end{subfigure}

  \caption{Posterior inclusion probabilities of the asymmetric three-body interaction indicators $c_{ijk}^{(a)}$.  
Each bar chart shows the inclusion probability for all ordered triplets $(i,j,k)$ with $i\neq j$, $j\neq k$, and $i\neq k$.  
The red horizontal line indicates the value 0.5, which is used as the cutoff threshold for constructing point estimates of the interaction structure.  
In all panels, the bar corresponding to the true nonzero interaction $\bar{c}_{123}^{(a)}$ is highlighted by a vertical blue dashed line.  
(a) Results for Configuration~1 (asynchronous dynamics).  
(b) Results for Configuration~2 (phase-locked dynamics).  
(c) Results for Configuration~3 (noise-perturbed phase-locked dynamics).
}
  \label{fig:cijk_a}
\end{figure*}
\begin{figure*}[t]
  \centering
  \begin{subfigure}{0.32\textwidth}
    \centering
    \caption{}
    \includegraphics[width=\linewidth]{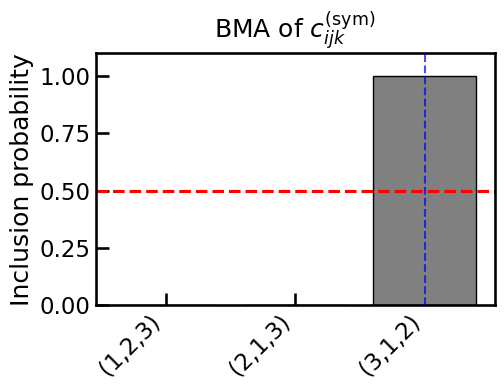}
    \label{fig:cijk_s_1}
  \end{subfigure}
  \hfill
  \begin{subfigure}{0.32\textwidth}
    \centering
    \caption{}
    \includegraphics[width=\linewidth]{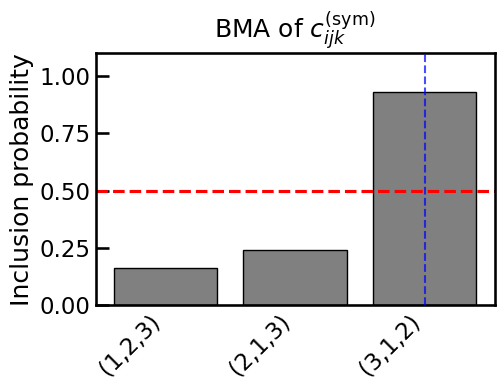}
    \label{fig:cijk_s_2}
  \end{subfigure}
  \hfill
  \begin{subfigure}{0.32\textwidth}
    \centering
    \caption{}
    \includegraphics[width=\linewidth]{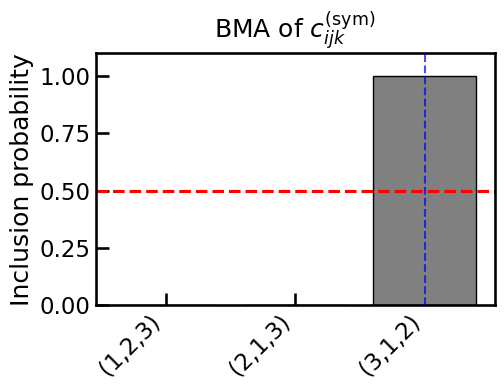}
    \label{fig:cijk_s_3}
  \end{subfigure}

  \caption{Posterior inclusion probabilities of the symmetric three-body indicators $c_{ijk}^{s}$. 
  Each bar chart shows the inclusion probability for all ordered triplets $(i,j,k)$ with $i\neq j$, $i\neq k$ and $j < k$.  
The red horizontal line indicates the value 0.5, which is used as the cutoff threshold for constructing point estimates of the interaction structure.  
In all panels, the bar corresponding to the true nonzero interaction $\bar{c}_{312}^{(s)}$ is highlighted by a vertical blue dashed line.  
(a) Results for Configuration~1 (asynchronous dynamics).  
(b) Results for Configuration~2 (phase-locked dynamics).  
(c) Results for Configuration~3 (noise-perturbed phase-locked dynamics).}
  \label{fig:cijk_s}
\end{figure*}
\begin{figure*}[t]
  \centering
  \begin{subfigure}{0.32\textwidth}
    \centering
    \caption{}
    \includegraphics[width=\linewidth]{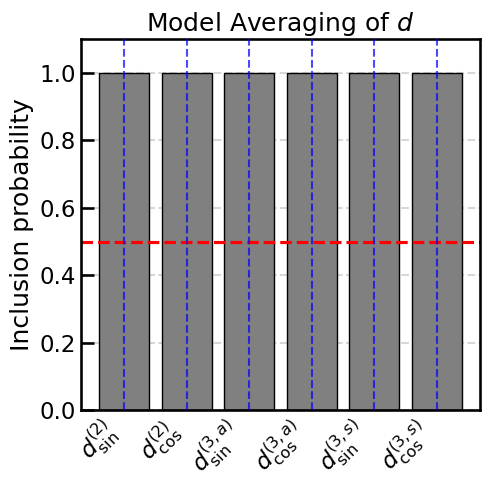}
    \label{fig:d_1}
  \end{subfigure}
  \hfill
  \begin{subfigure}{0.32\textwidth}
    \centering
    \caption{}
    \includegraphics[width=\linewidth]{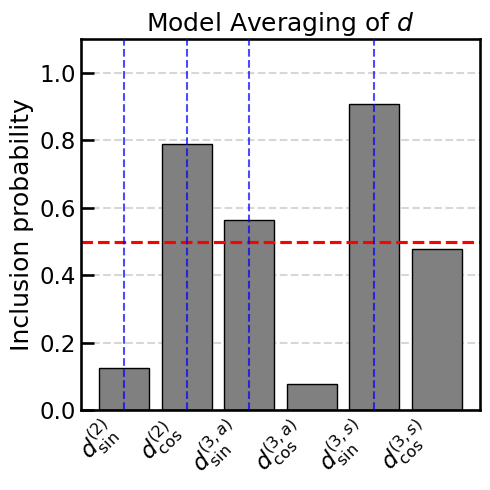}
    \label{fig:d_2}
  \end{subfigure}
  \hfill
  \begin{subfigure}{0.32\textwidth}
    \centering
    \caption{}
    \includegraphics[width=\linewidth]{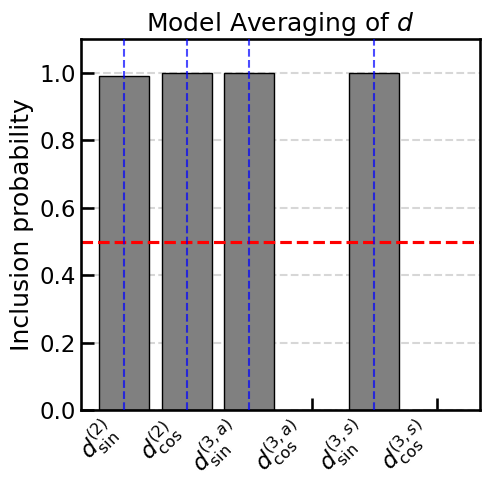}
    \label{fig:d_3}
  \end{subfigure}
  \caption{Posterior inclusion probabilities of the indicator vector $\bm d$ for the sine and cosine components associated with two-body interactions, asymmetric three-body interactions, and symmetric three-body interactions.  
The red horizontal line indicates the value 0.5, which is used as the cutoff threshold for constructing point estimates of the interaction structure. 
In all panels, the bar corresponding to true nonzero components are highlighted by a vertical blue dashed line.  
(a) Results for Configuration~1 (asynchronous dynamics).  
(b) Results for Configuration~2 (phase-locked dynamics).  
(c) Results for Configuration~3 (noise-perturbed phase-locked dynamics).
}
  \label{fig:d}
\end{figure*}
\begin{figure*}[t]
  \centering
  \begin{subfigure}{0.32\textwidth}
    \centering
    \caption{}
    \includegraphics[width=\linewidth]{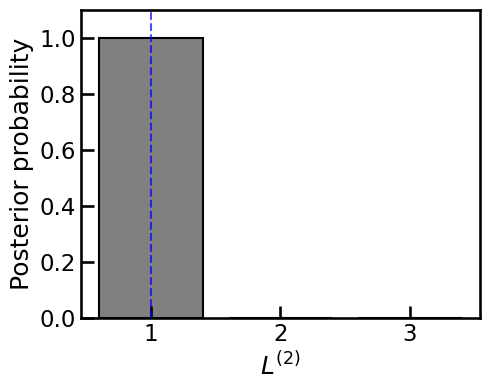}
    \label{fig:L2_1}
  \end{subfigure}
  \hfill
  \begin{subfigure}{0.32\textwidth}
    \centering
    \caption{}
    \includegraphics[width=\linewidth]{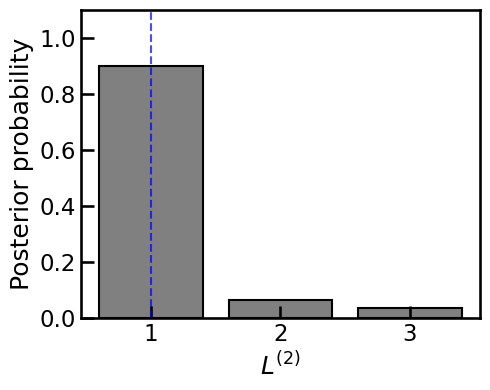}
    \label{fig:L2_2}
  \end{subfigure}
  \hfill
  \begin{subfigure}{0.32\textwidth}
    \centering
    \caption{}
    \includegraphics[width=\linewidth]{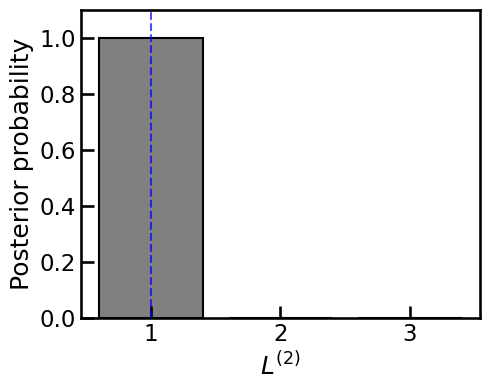}
    \label{fig:L2_3}
  \end{subfigure}
  \caption{Histograms of the marginal posterior distribution $P(L^{(2)} \mid \{\bm Y_i\}_{i=1}^N)$ sampled via parallel tempering (PT).  
The true harmonic order is $L^{(2)} = 1$ in all configurations.  
In all panels, the bar corresponding to true harmonic order is highlighted by a vertical blue dashed line. 
(a) Results for Configuration~1 (asynchronous dynamics).  
(b) Results for Configuration~2 (phase-locked dynamics).  
(c) Results for Configuration~3 (noise-perturbed phase-locked dynamics).
}
  \label{fig:L2}
\end{figure*}
\begin{figure*}[t]
  \centering
  \begin{subfigure}{0.32\textwidth}
    \centering
    \caption{}
    \includegraphics[width=\linewidth]{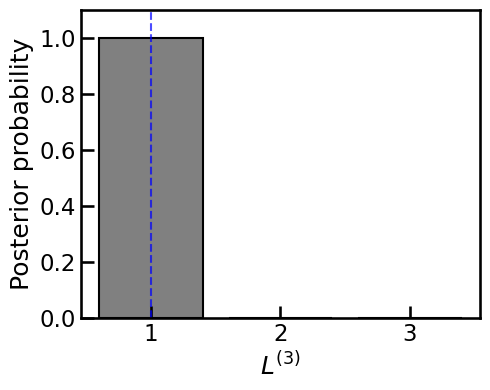}
    \label{fig:L3_1}
  \end{subfigure}
  \hfill
  \begin{subfigure}{0.32\textwidth}
    \centering
    \caption{}
    \includegraphics[width=\linewidth]{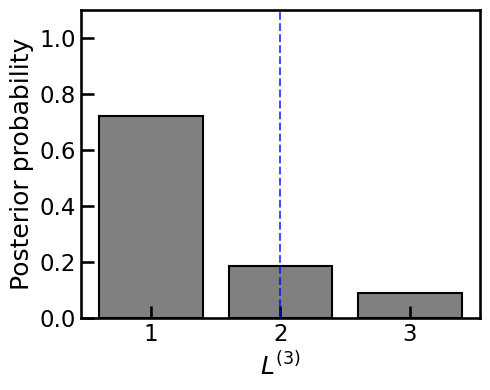}
    \label{fig:L3_2}
  \end{subfigure}
  \hfill
  \begin{subfigure}{0.32\textwidth}
    \centering
    \caption{}
    \includegraphics[width=\linewidth]{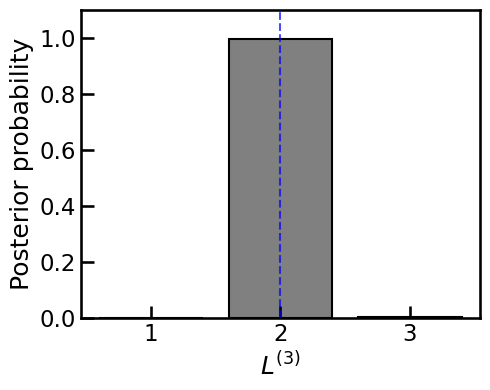}
    \label{fig:L3_3}
  \end{subfigure}
  \caption{Histograms of the marginal posterior distribution $P(L^{(3)} \mid \{\bm Y_i\}_{i=1}^N)$ sampled via parallel tempering (PT).  
The true harmonic order is $L^{(3)} = 1$ for Configuration 1 and $L^{(3)} = 2$ for Configuration 2. 
In all panels, the bar corresponding to true harmonic order is highlighted by a vertical blue dashed line. 
(a) Results for Configuration~1 (asynchronous dynamics).  
(b) Results for Configuration~2 (phase-locked dynamics).  
(c) Results for Configuration~3 (noise-perturbed phase-locked dynamics).
}
  \label{fig:L3}
\end{figure*}
\begin{figure*}[t]
  \centering
  \begin{subfigure}{0.48\textwidth}
    \centering
    \caption{}
    \includegraphics[width=\linewidth]{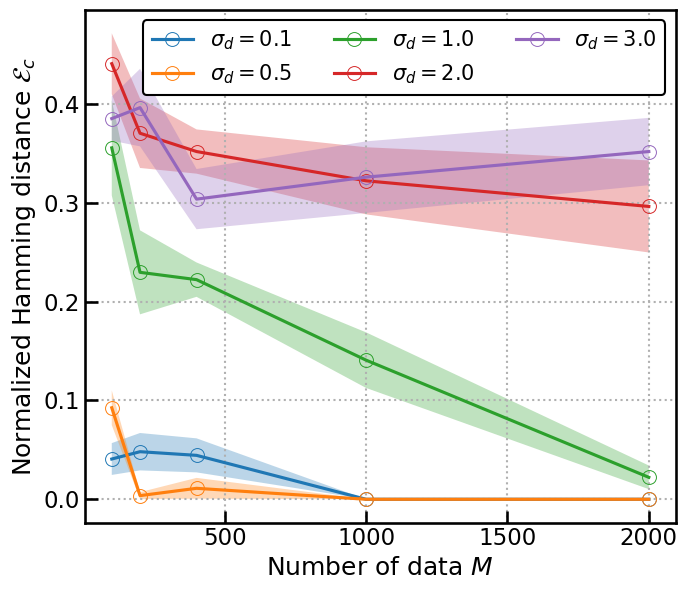}
    \label{fig:c_num_1}
  \end{subfigure}
  \hfill
  \begin{subfigure}{0.48\textwidth}
    \centering
    \caption{}
    \includegraphics[width=\linewidth]{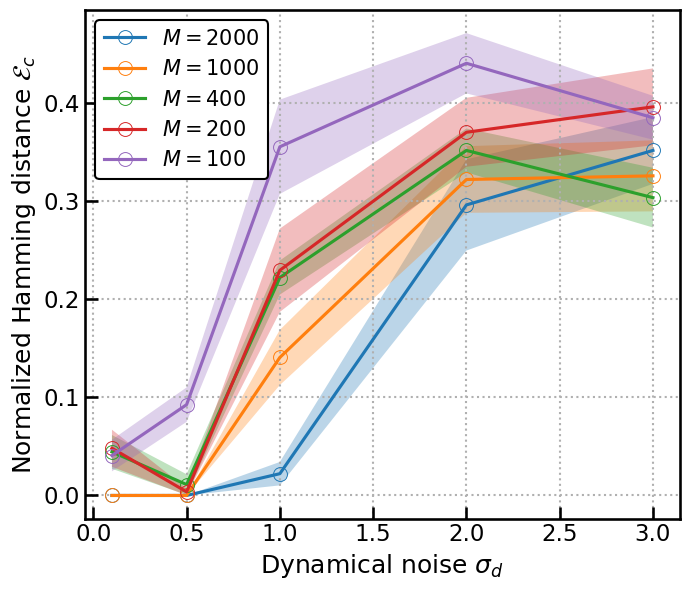}
    \label{fig:c_sig_1}
  \end{subfigure}
  \caption{Dependence of the Hamming distance $\mathcal E_c$ on (a) the number of data points and (b) the magnitude of dynamical noise.  
All experiments are performed under Configuration~1.  
Circles and shaded regions denote the mean and standard error computed over 10 independent runs, and solid lines connect the circles.
}
  \label{fig:c_1}
\end{figure*}
\begin{figure*}[t]
  \centering
  \begin{subfigure}{0.48\textwidth}
    \centering
    \caption{}
    \includegraphics[width=\linewidth]{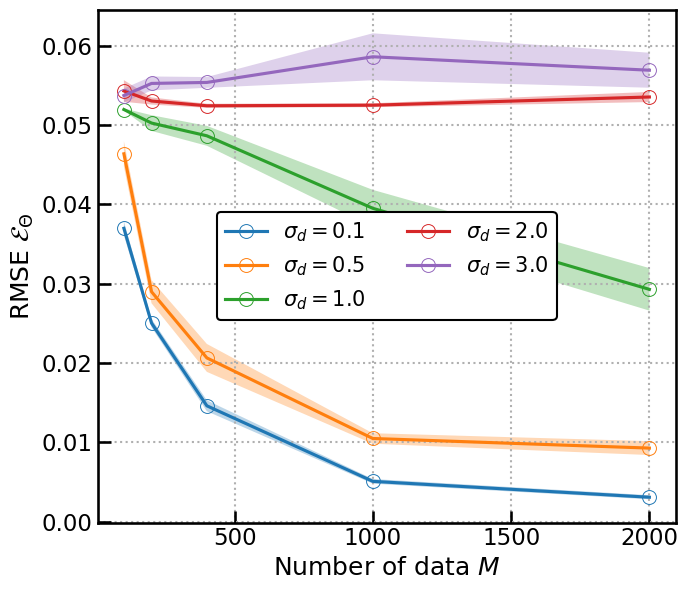}
    \label{fig:K_num_1}
  \end{subfigure}
  \hfill
  \begin{subfigure}{0.48\textwidth}
    \centering
    \caption{}
    \includegraphics[width=\linewidth]{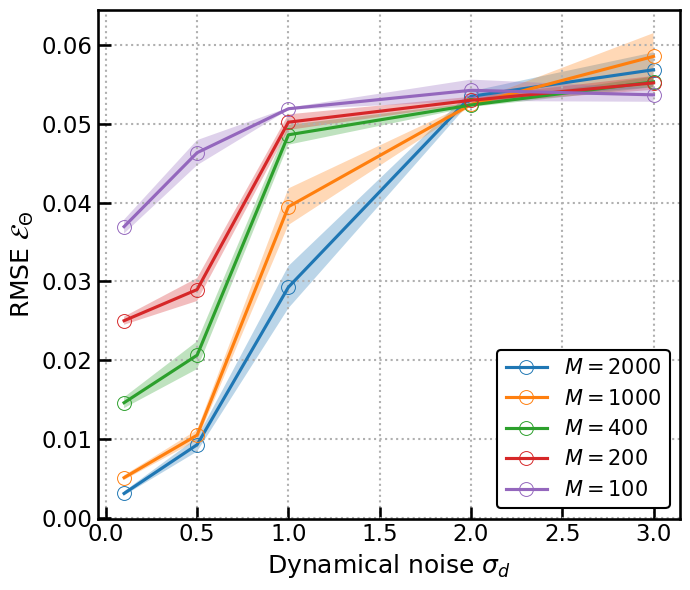}
    \label{fig:K_sig_1}
  \end{subfigure}

  \caption{Dependence of the MSE $\mathcal E_{\Theta}$ on (a) the number of data points and (b) the magnitude of dynamical noise.  
All experiments are performed under Configuration~1.  
Circles and shaded regions denote the mean and standard error computed over 10 independent runs, and solid lines connect the circles.}
  \label{fig:K_1}
\end{figure*}

\begin{figure*}[t]
  \centering
  \begin{subfigure}{0.48\textwidth}
    \centering
    \caption{}
    \includegraphics[width=\linewidth]{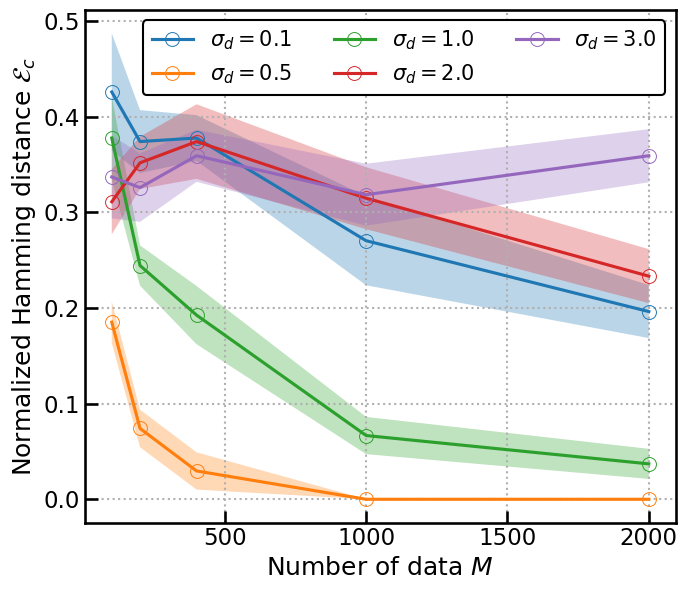}
    \label{fig:c_num_2}
  \end{subfigure}
  \hfill
  \begin{subfigure}{0.48\textwidth}
    \centering
    \caption{}
    \includegraphics[width=\linewidth]{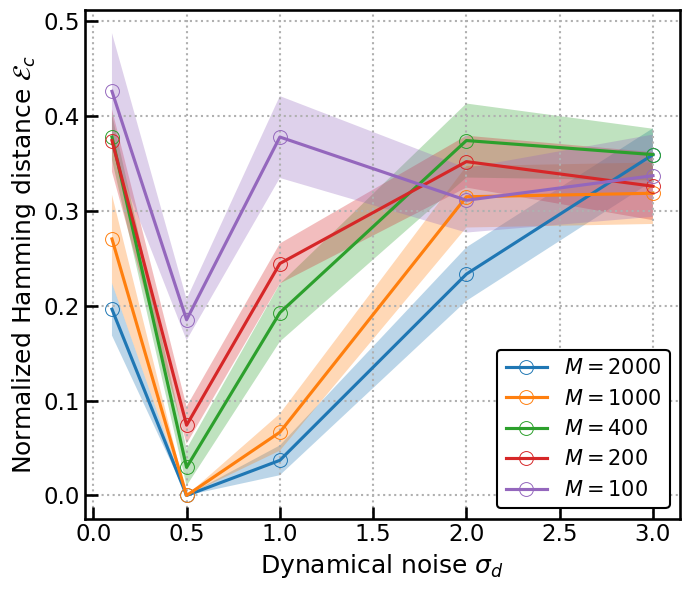}
    \label{fig:c_sig_2}
  \end{subfigure}
  \caption{Dependence of the Hamming distance $\mathcal E_c$ on (a) the number of data points and (b) the magnitude of dynamical noise.  
All experiments are performed under Configuration~2.  
Circles and shaded regions denote the mean and standard error computed over 10 independent runs, and solid lines connect the circles.}
  \label{fig:c_2}
\end{figure*}
\begin{figure*}[t]
  \centering
  \begin{subfigure}{0.48\textwidth}
    \centering
    \caption{}
    \includegraphics[width=\linewidth]{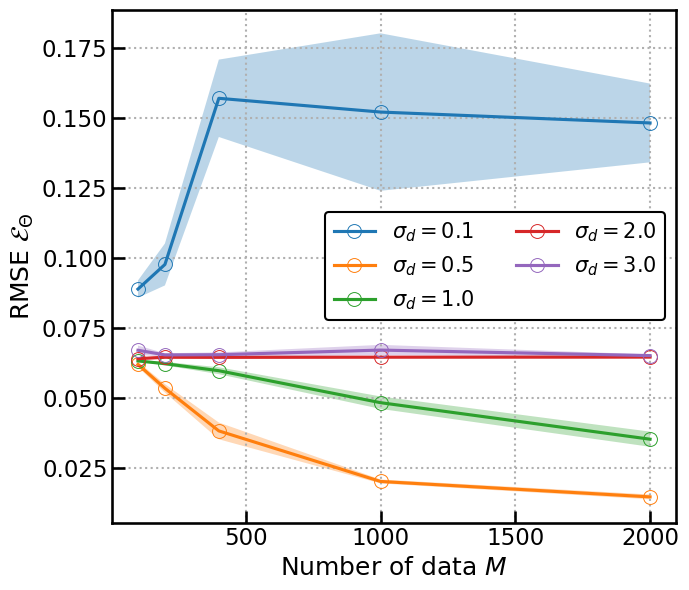}
    \label{fig:K_num_2}
  \end{subfigure}
  \hfill
  \begin{subfigure}{0.48\textwidth}
    \centering
    \caption{}
    \includegraphics[width=\linewidth]{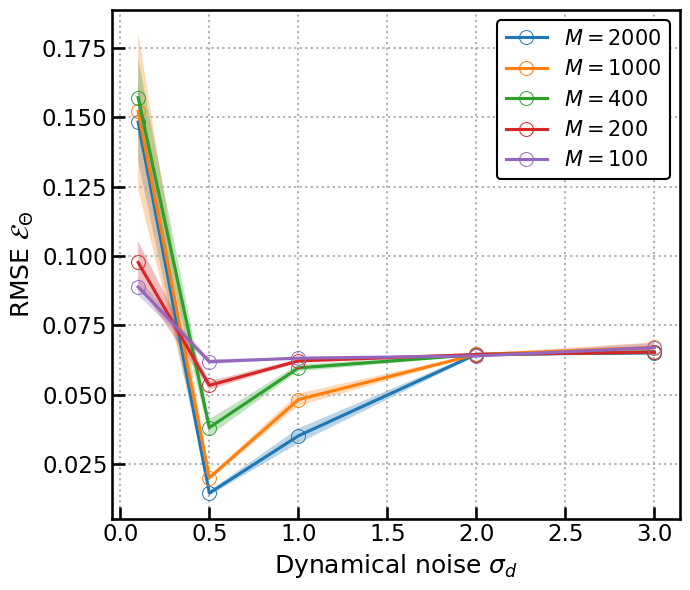}
    \label{fig:K_sig_2}
  \end{subfigure}

  \caption{Dependence of the MSE $\mathcal E_{\Theta}$ on (a) the number of data points and (b) the magnitude of dynamical noise.  
All experiments are performed under Configuration~2.  
Circles and shaded regions denote the mean and standard error computed over 10 independent runs, and solid lines connect the circles.}
  \label{fig:K_2}
\end{figure*}
\begin{figure}[t]
  \centering

  \begin{subfigure}{0.48\linewidth}
    \centering
    \caption{}
    \includegraphics[width=\linewidth]{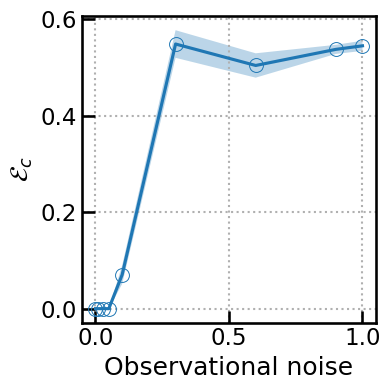}
    \label{fig:obs_1}
  \end{subfigure}
  \hfill
  \begin{subfigure}{0.48\linewidth}
    \centering
    \caption{}
    \includegraphics[width=\linewidth]{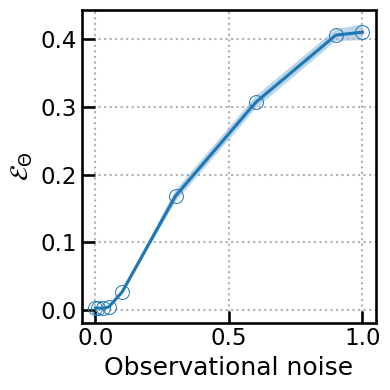}
    \label{fig:obs_2}
  \end{subfigure}

  \vspace{0.25cm}

  \begin{subfigure}{0.48\linewidth}
    \centering
    \caption{}
    \includegraphics[width=\linewidth]{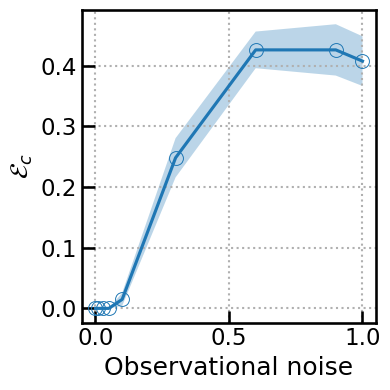}
    \label{fig:obs_3}
  \end{subfigure}
  \hfill
  \begin{subfigure}{0.48\linewidth}
    \centering
    \caption{}
    \includegraphics[width=\linewidth]{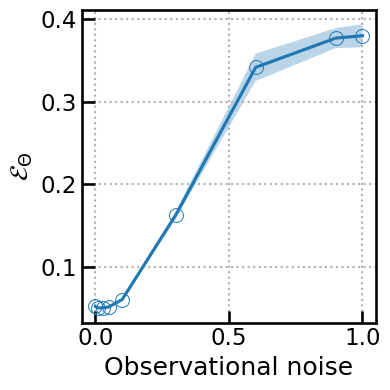}
    \label{fig:obs_4}
  \end{subfigure}

  \caption{(a) Dependence of the Hamming distance $\mathcal{E}_c$ on the magnitude of observational noise for Configuration~1.  
(b) Dependence of the MSE $\mathcal{E}_{\Theta}$ on the magnitude of observational noise for Configuration~1.  
(c) Dependence of the Hamming distance $\mathcal{E}_c$ on the magnitude of observational noise for Configuration~2.  
(d) Dependence of the MSE $\mathcal{E}_{\Theta}$ on the magnitude of observational noise for Configuration~2.  
For all panels, circles and shaded regions denote the mean and standard error computed over 10 independent runs, and solid lines connect the circles.
}
  \label{fig:obs_2x2}
\end{figure}
\section{Numerical experiments}\label{sec5}
In this section, we demonstrate the effectiveness of the proposed Bayesian sparse modeling framework through numerical experiments. 
We focus on oscillatory networks and aim to recover the underlying interaction structure from time–series phase observations.
Synthetic datasets are generated by specifying ground–truth network structures and coupling strengths, and comparing them with the estimates obtained by our method.

\paragraph{Evaluation metrics.}
For each oscillator $i$, the true indicator vector $\bar{\bm c}_i$ and the
estimated indicator vector $\hat{\bm c}_i$ are compared using the average
$\ell_1$ error:
\begin{equation}
\mathcal E_c = \frac{1}{N\Gamma}\sum_{i=1}^N 
\left\lVert\, \bar{\bm c}_i - \hat{\bm c}_i \,\right\rVert_1 ,
\end{equation}
which measures the discrepancy between the true and inferred interaction
structures.  
This metric is equivalent to the normalized Hamming distance between the
binary indicator vectors.

To evaluate the accuracy of coupling strength estimation, we note that the posterior mean 
$\hat{\bm\Theta}_i^c \in \mathbb R^{|\hat{\bm c}_i|_0}$ is obtained only on the active subspace with dimension $|\hat{\bm c}_i|_0$.
Thus, for comparison with the ground truth $\bar{\bm\Theta}_i\odot \bar{\bm c}_i\in\mathbb{R}^{\Gamma}$,
we construct a $\Gamma$–dimensional estimate by padding zeros for inactive components:
\[
\hat{\bm\Theta}_i^{(\Gamma)} 
= \mathrm{pad}\bigl(\hat{\bm\Theta}_i^c,\, \hat{\bm c}_i \bigr),
\]
where $\mathrm{pad}(\cdot)$ places each entry of $\hat{\bm\Theta}_i^c$ in the position corresponding to the active components of $\hat{\bm c}_i$ and fills the other entries with zeros.
The estimation error for coupling strengths is then quantified by
\begin{equation}
\mathcal E_{\Theta} = \sqrt{\frac{1}{N\Gamma}\sum_{i=1}^N 
\left\lVert\, 
  \bar{\bm\Theta}_i\odot\bar{\bm c}_i
  - \hat{\bm\Theta}_i^{(\Gamma)}
\,\right\rVert_2^2 }.
\end{equation}
This metric, which is called root mean square error (RMSE), compares both the magnitude of inferred couplings and their alignment with the true active components.

\subsection{Oscillatory network}

We apply the proposed Bayesian sparse modeling framework to generalized oscillator networks described by
\begin{align}
\frac{dX_i}{dt}
 &= \omega_i 
   + \sum_{j\neq i}\sum_{l=1}^{L^{(2)}}
       K_{ij}^{(l)}\,\sin\!\bigl(l(X_j-X_i)+\alpha_{ij}^{(l)}\bigr) \nonumber \\
  &\quad + \sum_{\substack{j\neq i,\,k\neq i \\ j\neq k}}
       \sum_{l=1}^{L^{(3)}}
       K_{ijk}^{(l)}\,\sin\!\bigl(l(2X_k-X_i-X_j)+\alpha_{ijk}^{(l)}\bigr) \nonumber \\
  &\quad + \sum_{\substack{j\neq i,\,k\neq i \\ j<k}}
       \sum_{l=1}^{L^{(3)}}
       K_{ijk}^{(l)\prime}\,\sin\!\bigl(l(X_k+X_j-2X_i)+\alpha_{ijk}^{(l)'}\bigr),
\label{eq:oscillator_target}
\end{align}
where $\omega_i$ is the intrinsic frequency, and the second and third terms represent pairwise, asymmetric three--body, and symmetric three--body interactions.  
Our goal is to infer the presence of each of these interactions and to determine the corresponding harmonic and phase--lag structure.

\paragraph{Basis construction.}
To map this model into the general formalism of Eq. (\ref{general_formulation}),
we express all interaction terms using sine and cosine bases.
The cosine bases represent the effect of the phase lag.
For each ordered pair $(i,j)$ with $i\neq j$ we consider
\[
\mathcal{B}_{ij}^{(2)}
  = \{\sin(l\Delta X_{ij}),\;\cos(l\Delta X_{ij})\}_{l=1}^{L^{(2)}},
\]
where $\Delta X_{ij} = X_j - X_i$.
For three--body interactions we distinguish two sets of bases:
\[
\mathcal{B}_{ijk}^{(3,\mathrm{a})}
  = \{\sin(l\phi^{(\mathrm{a})}_{ijk}),\;\cos(l\phi^{(\mathrm{a})}_{ijk})\}_{l=1}^{L^{(3)}},
\]
\[
\mathcal{B}_{ijk}^{(3,\mathrm{s})}
  = \{\sin(l\phi^{(\mathrm{s})}_{ijk}),\;\cos(l\phi^{(\mathrm{s})}_{ijk})\}_{l=1}^{L^{(3)}},
\]
with
\[
\phi^{(\mathrm{a})}_{ijk}=2X_k-X_i-X_j,
\qquad
\phi^{(\mathrm{s})}_{ijk}=X_k+X_j-2X_i,
\]
where $i$, $j$, and $k$ are mutually distinct.
Thus asymmetric and symmetric three--body interactions are treated as
completely separate candidate components.

The set of basis functions $\bm g_{i,m}$ is constructed by concatenating
all bases in $\mathcal{B}_{ij}^{(2)}$, $\mathcal{B}_{ijk}^{(3,\mathrm{a})}$,
and $\mathcal{B}_{ijk}^{(3,\mathrm{s})}$ for fixed $i$ and time index $m$.
By additionally including the constant basis \(1\) to represent the intrinsic
frequency, the total number of candidate basis functions is $\Gamma = 1 + 2L^{(2)}(N-1)+3L^{(3)}(N-1)(N-2)$.

\paragraph{Indicators and vectorization.}
For each interaction class introduced in Sec.~III, we assign a binary
structural indicator to every basis function.
Specifically,
(i) $c_{ij}^{(2)}$ denotes whether the pairwise interaction
$\mathcal{B}_{ij}^{(2)}$ is active,
(ii) $c_{ijk}^{(3,\mathrm a)}$ denotes whether the asymmetric three--body basis
$\mathcal{B}_{ijk}^{(3,\mathrm a)}$ is active,
and (iii) $c_{ijk}^{(3,\mathrm s)}$ denotes whether the symmetric three--body
basis $\mathcal{B}_{ijk}^{(3,\mathrm s)}$ is active.

To independently control the presence of sine and cosine components
(i.e., phase--lag effects), we introduce a six-dimensional binary
indicator vector
\[
\bm d
=
(d_{\sin}^{(2)}, d_{\cos}^{(2)},
 d_{\sin}^{(3,\mathrm a)}, d_{\cos}^{(3,\mathrm a)},
 d_{\sin}^{(3,\mathrm s)}, d_{\cos}^{(3,\mathrm s)})^\top
\in\{0,1\}^6.
\]
The cosine components represent phase--lag contributions, so setting
$d_{\cos}^{(\cdot)}=0$ suppresses all phase--lag terms in the
corresponding interaction class.

For oscillator $i$, the full set of interaction-dependent basis functions
is therefore
\[
\begin{aligned}
&\mathcal{B}_i^{\mathrm{(int)}}
={}\;
\Bigl\{
  c_{ij}^{(2)} d_{\sin}^{(2)} \sin(l\Delta X_{ij}),\;
  c_{ij}^{(2)} d_{\cos}^{(2)} \cos(l\Delta X_{ij})
\Bigr\}_{j\neq i,\; l=1}^{L^{(2)}} \\
&{}\cup
\Bigl\{
  c_{ijk}^{(3,\mathrm a)} d_{\sin}^{(3,\mathrm a)} \sin(l\phi^{(\mathrm a)}_{ijk}),\;
  c_{ijk}^{(3,\mathrm a)} d_{\cos}^{(3,\mathrm a)} \cos(l\phi^{(\mathrm a)}_{ijk})
\Bigr\}_{\substack{i,j,k\;\mathrm{dist.}\\ l=1}}^{L^{(3)}} \\
&{}\cup
\Bigl\{
  c_{ijk}^{(3,\mathrm s)} d_{\sin}^{(3,\mathrm s)} \sin(l\phi^{(\mathrm s)}_{ijk}),\;
  c_{ijk}^{(3,\mathrm s)} d_{\cos}^{(3,\mathrm s)} \cos(l\phi^{(\mathrm s)}_{ijk})
\Bigr\}_{\substack{i,j,k\;\mathrm{dist.}\\ l=1}}^{L^{(3)}} .
\end{aligned}
\]
These three blocks correspond exactly to the basis families
$\mathcal{B}_{ij}^{(2)}$, $\mathcal{B}_{ijk}^{(3,\mathrm a)}$,
and $\mathcal{B}_{ijk}^{(3,\mathrm s)}$ defined in the previous subsection.

Finally, each oscillator also includes one intrinsic basis function
$g_{i,m}^{(0)} = 1$,
with associated indicator $c_i^{(0)}=1$ and parameter
$\Theta_i^{(0)} = \omega_i$.
Since this term is unrelated to the sine and cosine bases, there is no need to introduce an additional indicator for it. We therefore fix $d_{i,0}=1$.

The complete set of basis functions for oscillator $i$ is thus
\[
\mathcal{B}_i = \{ g_{i,m}^{(0)} \} \cup \mathcal{B}_i^{\mathrm{(int)}}.
\]

All indicators are vectorized into a single effective indicator vector
\[
\tilde{\bm c}_i
= \{\tilde c_{i}^{(\gamma)} = c_{i}^{(\gamma)} d_{i}^{(\gamma)}\}_{\gamma \in \Gamma},
\]
where $\Gamma$ is the index set of all basis functions
(including pairwise, asymmetric and symmetric three--body,
all harmonics, and the intrinsic term).
This effective indicator simultaneously determines  
(i) whether an interaction is present,  
(ii) whether it is pairwise or three--body,  
(iii) whether it is asymmetric or symmetric,  
(iv) which harmonic components are included, and  
(v) whether the phase--lag (cosine) components are active.

\paragraph{Compact representation.}
Let $\Theta_{i,\gamma}$ denote the coefficient of basis $g_{i,\gamma}$.
The discrete-time dynamics become
\begin{equation}
\Delta X_{im}
  = \Delta t \sum_{\gamma\in\Gamma}
      \tilde{c}_{i}^{(\gamma)}\,
      \Theta_{i}^{\gamma}\,
      g_{i,m}^{(\gamma)},
\end{equation}
which can be written in vector form as
\begin{equation}
\Delta X_{im}
  = \Delta t\,(\tilde{\bm c}_i \odot \bm\Theta_i)^\mathsf{T}\bm g_i^{(m)},
\end{equation}
matching the general sparse linear representation Eq. (\ref{general_formulation}).

Bayesian model averaging over the binary indicators
$c_{ij}^{(2)}$, $c_{ijk}^{(3,\mathrm a)}$, $c_{ijk}^{(3,\mathrm s)}$
and the phase–lag indicators $d_{i,\gamma}$
provides posterior inclusion probabilities for all candidate interactions,
yielding uncertainty-aware estimates of the network structure and
phase–lag components.
In addition, the harmonic orders $L^{(2)}$ and $L^{(3)}$ are sampled as
discrete model hyperparameters, and their posterior distributions
directly quantify the harmonic richness without model averaging.

\begin{table}[t]
  \centering
  \small
  \caption{Prior distributions and their hyperparameters.}
  \label{tab:priors}
  \begin{tabular}{lll}
    \hline
    Quantity & Prior & Hyperparameters \\
    \hline
    $c_i^{(\gamma)}$ 
      & $\mathrm{Bernoulli}(p)$ 
      & $p=0.5$ \\[2pt]

    $\sigma_i$ 
      & $\mathrm{Unif}(\sigma_{\min}, \sigma_{\max})$ 
      & $\sigma_{\min}=0.025, \sigma_{\max}=5.77$ \\[2pt]

    $\tau_{i\gamma}$ 
      & $\mathrm{Unif}(\tau_{\min}, \tau_{\max})$ 
      & $\tau_{\min}=0.01, \tau_{\max}=10.0$ \\[2pt]

    $L^{(2)}$ 
      & $\mathrm{Unif}\{1,\dots,L^{(2)}_{\max}\}$ 
      & $L^{(2)}_{\max}=3$ \\[2pt]

    $L^{(3)}$ 
      & $\mathrm{Unif}\{1,\dots,L^{(3)}_{\max}\}$ 
      & $L^{(3)}_{\max}=3$ \\
    \hline
  \end{tabular}
\end{table}

For the prior distributions, all hyperparameter settings are summarized in Table~\ref{tab:priors}.
For parallel tempering, we use $R = 40$ replicas, and the temperature ladder is defined by a geometric factor $\eta = 1.3$. 
The code used to reproduce the results in this study is available at:
https://github.com/shuhei-desu/bayes-spm-dynamical-system.

\subsubsection{Configuration 1: Asynchronous dynamics}

We first demonstrate our framework by inferring the interaction structure
of a small oscillatory network.
We consider a system of three oscillators with the true sparse interactions
\[
c_{21}=1,\; c_{31}=1,\; c_{123}^{(\mathrm a)}=1,\; c_{312}^{(\mathrm s)}=1,
\]
and all other indicators set to zero.
The corresponding coupling strengths are
\[
K_{21}=K_{31}=K_{123}^{(\mathrm a)}=K_{312}^{(\mathrm s)}=0.5 .
\]
The phase lags are set to
\[
\alpha_{ij}=1.0,\qquad 
\alpha_{ijk}=1.0,\qquad 
\alpha_{ijk}'=1.0,
\]
and the interaction orders to \(L^{(2)}=1\) and \(L^{(3)}=1\).
The intrinsic frequencies are
$(\omega_1, \omega_2, \omega_3)^T=(0.5, 1.0, 1.5)^{\mathsf T}$,
with initial condition of phases \((X_{1,0}, X_{2,0}, X_{3,0})=(0.0, 2.0, 4.0)\).
This configuration yields an asynchronous trajectory without phase locking.
We use \(M=2000\) data points and dynamical noise \(\sigma_d=0.1\);
observational noise is removed (\(\sigma_o=0\)) to isolate the effect of
dynamical noise.
The generated time series is shown in Fig.~\ref{fig:phase1}.

The posterior inclusion probabilities of pairwise indicators \(c_{ij}\), asymmetric three-body indicators \(c_{ijk}^{(\mathrm a)}\), symmetric three-body indicators \(c_{ijk}^{(\mathrm s)}\), and sine/cosine indicators \(d\) are shown in Figs.~\ref{fig:cij1}, \ref{fig:cijk_a_1}, \ref{fig:cijk_s_1}, and \ref{fig:d_1}, respectively.
Figs.~\ref{fig:cij1}--\ref{fig:cijk_s_1} demonstrate that the proposed Bayesian model averaging (BMA) assigns high posterior probability to all components included in the true network and low probability to irrelevant ones.
Figure~\ref{fig:d_1} shows that both sine and cosine bases, corresponding to nonzero phase lags, are correctly identified as essential.

The posterior distributions of the maximum harmonic orders \(L^{(2)}\) and \(L^{(3)}\) are shown in Figs.~\ref{fig:L2_1} and \ref{fig:L3_1}.
Both orders are recovered with high confidence, confirming that the method successfully infers the correct harmonic complexity.

\subsubsection{Configuration 2: Phase-locked dynamics}
Next, we consider a configuration similar to Configuration~1, but with
modified phase-lag and higher-harmonic settings.
We again set
\[
c_{21}=1,\; c_{31}=1,\; c_{123}^{(\mathrm a)}=1,\; c_{312}^{(\mathrm s)}=1,
\]
and all other indicators to zero, while keeping the same coupling strengths as
before,
\[
K_{21}=K_{31}=K_{123}^{(\mathrm a)}=K_{312}^{(\mathrm s)}=0.5 .
\]
The phase lags and interaction order are changed to
\[
\alpha_{ij}=1.0,\quad
\alpha_{ijk}=0.0,\quad
\alpha_{ijk}'=0.0,\quad
L^{(2)}=1,\quad
L^{(3)}=2,
\]
and intrinsic frequencies to $(\omega_1, \omega_2, \omega_3)^{\mathsf T}
=(0.4, 0.8, 1.2)$ with initial phases
\((X_{1,0}, X_{2,0}, X_{3,0})=(0.0, 2.0, 4.0)\).
We again use \(M=2000\) data points and \(\sigma_d=0.1\), \(\sigma_o=0\).
The resulting trajectory (Fig.~\ref{fig:phase2}) exhibits a clear phase-locked state with nearly constant phase differences.

The inferred posterior inclusion probabilities for \(c_{ij}\), \(c_{ijk}^{(\mathrm a)}\), and \(c_{ijk}^{(\mathrm s)}\) are shown in
Figs.~\ref{fig:cij2}, \ref{fig:cijk_a_2}, and \ref{fig:cijk_s_2}.
Incorrect interactions such as \(c_{12}\), \(c_{13}\), and \(c_{321}^{(\mathrm a)}\) are assigned high posterior probability, and many irrelevant connections receive nonzero inclusion probability.
Figure~\ref{fig:d_2} shows that the essential sine basis for pairwise interactions receives low probability, while non-essential cosine bases for symmetric three-body components receive undesirably high probability.

The posterior distributions of \(L^{(2)}\) and \(L^{(3)}\) (Figs.~\ref{fig:L2_2} and \ref{fig:L3_2}) further illustrate the failure: in particular, the inferred three-body order \(L^{(3)}\) is completely incorrect.

These failures arise from the phase-locked nature of the trajectory.
With constant phase differences, the basis functions become effectively linearly dependent, destroying identifiability of the true interaction structure and harmonic orders.

\subsubsection{Configuration 3: Breaking phase locking with larger dynamical noise}

To examine the effect of dynamical noise in phase-locked regimes, we repeat Configuration~2 with an increased noise level $\sigma_d = 0.5$, while keeping all other settings unchanged.
The resulting trajectory (Fig.~\ref{fig:phase3}) exhibits fluctuations, producing richer dynamical variability.

The inferred inclusion probabilities are shown in
Figs.~\ref{fig:cij3}, \ref{fig:cijk_a_3}, \ref{fig:cijk_s_3}, \ref{fig:d_3}, \ref{fig:L2_3}, and \ref{fig:L3_3}.
In contrast to Configuration~2, all true interactions, phase-lag components, and harmonic orders are recovered with high credibility.
The increase in dynamical variability resolves the effective linear dependence of the bases, restoring identifiability.

\subsubsection{Robustness analysis}

To evaluate the robustness of our method with respect to the number of data points and the magnitude of dynamical noise, we repeated data generation and estimation for various settings and computed the mean and standard error of the metrics \(\mathcal E_c\) and \(\mathcal E_{\Theta}\).
The results are summarized in Figs.~\ref{fig:c_num_1}--\ref{fig:c_sig_2}.

\paragraph{Asynchronous regime (Configuration 1).}
We first adopt the same setting as Configuration~1 and vary the number of data points and the variance of dynamical noise.
Figures~\ref{fig:c_num_1} and \ref{fig:c_sig_1} show $\mathcal E_c$ as a function of the number of data points and dynamical noise, respectively, and Figs.~\ref{fig:K_num_1} and \ref{fig:K_sig_1} show the corresponding results for $\mathcal E_{\Theta}$.
As expected for an asynchronous trajectory that explores the phase space sufficiently, the dependence on the number of samples is essentially monotonic: both $\mathcal E_c$ and $\mathcal E_{\Theta}$ decrease as more data become available.
The effect of dynamical noise is comparatively mild. 
Even at a small noise level (e.g., $\sigma_d = 0.1$), both the network structure and the coupling strengths are already recovered with reasonably high accuracy, and introducing moderate noise further improves the performance only slightly.
The errors attain their minimum around $\sigma_d \approx 0.5$, but the gain over $\sigma_d = 0.1$ is modest.
For larger noise, the errors gradually increase as the dynamics are dominated by fluctuations.
Thus, in the asynchronous regime, dynamical noise plays a secondary role: it can slightly enhance identifiability by diversifying the observed trajectories, but accurate reconstruction is already possible in the low-noise limit.

\paragraph{Phase-locked regime (Configurations 2 and 3).}
We next consider the phase-locked regime and adopt the settings of Configurations~2 and~3, again varying the number of data points and the variance of dynamical noise.
Figures~\ref{fig:c_num_2} and \ref{fig:c_sig_2} show the resulting $\mathcal E_c$, while Figs.~\ref{fig:K_num_2} and \ref{fig:K_sig_2} show $\mathcal E_{\Theta}$.
As in the asynchronous case, increasing the number of data points monotonically improves estimation.
In contrast, the dependence on dynamical noise is now much more pronounced.
When the noise is small (Configuration~2, $\sigma_d = 0.1$), the system remains close to a perfectly phase-locked state, many basis functions become effectively linearly dependent, and both $\mathcal E_c$ and $\mathcal E_{\Theta}$ remain large: the true interaction structure is not reliably recovered.
Introducing moderate dynamical noise (Configuration~3, $\sigma_d \approx 0.5$) breaks the strict phase locking, generates richer phase fluctuations, and dramatically improves identifiability of both the network structure and the coupling strengths.
For larger noise, however, the signal is again overwhelmed by fluctuations and the errors increase.
Compared with the asynchronous regime, the improvement around $\sigma_d \approx 0.5$ is therefore much more substantial in the phase-locked regime, indicating that an intermediate level of dynamical noise is essential to resolve degeneracies in the basis functions and to maximize identifiability of the true interaction structure.

\subsubsection{Effect of observational noise}
We next examine the impact of observational noise on estimation performance.
To isolate this effect, we focus on two configurations, namely Configuration~1 and Configuration~3, for which the true network structure and interaction orders are accurately recovered in the absence of observational noise.

Figures~\ref{fig:obs_1} and~\ref{fig:obs_2} show the dependence of 
$\mathcal{E}_c$ and $\mathcal{E}_{\Theta}$ on the magnitude of observational noise for Configuration~1.
Similarly, Figs.~\ref{fig:obs_3} and~\ref{fig:obs_4} present the corresponding results for Configuration~3.

In both cases, the estimation accuracy deteriorates as observational noise increases.
Nevertheless, for small noise levels (up to approximately $0.1$), both the network structure and coupling strengths remain reliably identifiable, demonstrating that the proposed Bayesian framework is robust to small observational noise.

\subsection{Data-driven phase reduction for a metronome system}

\begin{figure*}[t]
  \centering
  \includegraphics[width=\linewidth]{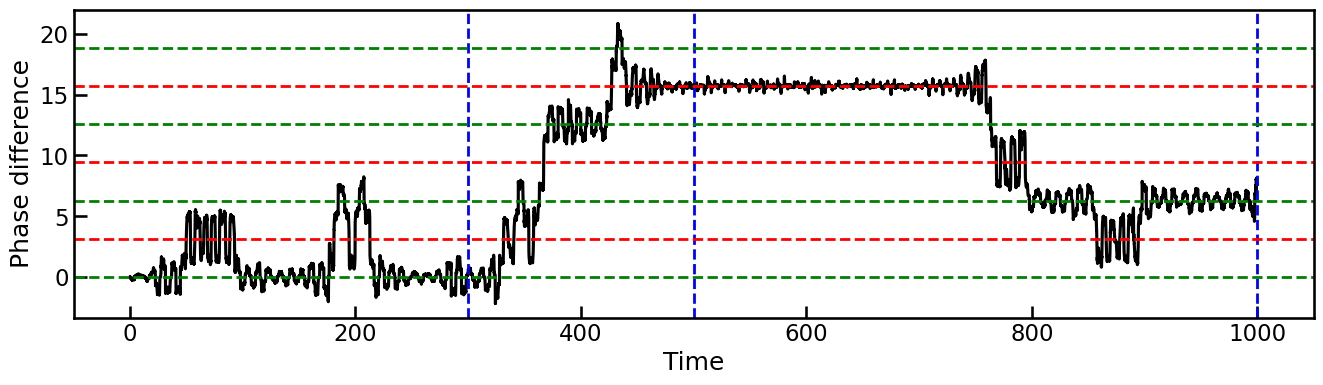}
  \caption{Phase difference dynamics extracted from simulations of two coupled metronomes.
The trajectory shows the temporal evolution of the phase difference 
$\psi = \theta_1 - \theta_2$.
Green dashed lines indicate in-phase synchronization states 
($\psi = 2m\pi$), while red dashed lines indicate anti-phase synchronization states 
($\psi = (2m+1)\pi/2$), where $m$ is an integer.
The three vertical blue dashed lines correspond to the boundaries of the
time windows used for the analysis: $[0,300]$, $[0,500]$, and $[0,1000]$.}
  \label{fig:phase_diff}
\end{figure*}

\begin{figure*}[t]
  \centering
  \begin{subfigure}{0.32\textwidth}
    \centering
    \caption{}
    \includegraphics[width=\linewidth]{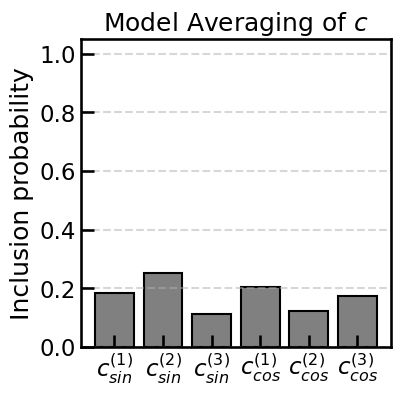}
    \label{fig:d_1_metro}
  \end{subfigure}
  \hfill
  \begin{subfigure}{0.32\textwidth}
    \centering
    \caption{}
    \includegraphics[width=\linewidth]{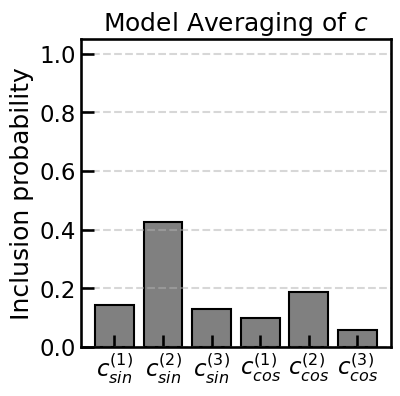}
    \label{fig:d_2_metro}
  \end{subfigure}
  \hfill
  \begin{subfigure}{0.32\textwidth}
    \centering
    \caption{}
    \includegraphics[width=\linewidth]{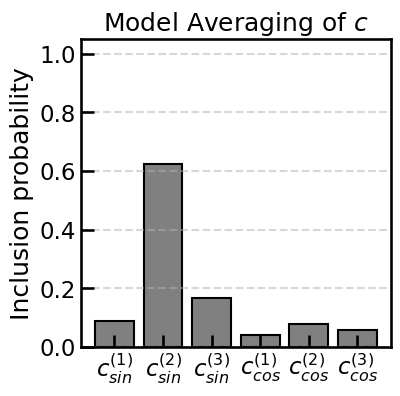}
    \label{fig:d_3_metro}
  \end{subfigure}
  \caption{Posterior inclusion probabilities of the indicator vector for the sine and cosine components.
Each panel shows the results of the analysis using different time windows:
(a) $[0,300]$, 
(b) $[0,500]$, 
(c) $[0,1000]$.
}
  \label{fig:d_metro}
\end{figure*}

In the previous section, we considered the setting in which the true dynamical model is contained within the candidate model class. In realistic situations, however, the governing equations are typically unknown and may not belong to the assumed model family. To examine how the proposed method behaves in such circumstances, we investigate a mismatched setting in which the statistical model does not include the true dynamics.

In this section, we perform a data driven phase reduction of a metronome system. A coupled metronome on a movable platform is a mechanically simple yet dynamically rich oscillator system: its full dynamics are described by underdamped nonlinear equations involving escapement mechanisms, damping, and coupling. The system exhibits diverse nonlinear phenomena including bistability of in-phase and anti-phase synchronization~\cite{wu2012anti}, oscillation quenching~\cite{goldsztein2022coupled,kato2024weakly}, chaos~\cite{ulrichs2009synchronization}, and complex transient dynamics, making it a suitable testbed for identifying effective dynamical representations. By fitting an reduced phase model to trajectories generated from the mechanical equations, we evaluate whether the proposed framework can select the functional components required to describe the motion and provide uncertainty aware data driven phase reduction.

In Ref.~\cite{kato2024weakly} the authors proposed an analytically tractable model of two identical metronomes 
placed on a common movable platform are described by the following nondimensional equations of motion for the relative displacements $x_i$ of each oscillator measured from the platform:
\begin{align}
\ddot x_i &= - x_i - \epsilon\left[\mu(x_1 + x_2) + \beta \dot x_i - g(x_i,\dot x_i) \right] \nonumber \\
&+ \epsilon^2 \mu \left[ -\beta(\dot x_1 + \dot x_2) + g(x_1,\dot x_1) + g(x_2,\dot x_2) \right].
\end{align}
The parameter $\beta$ represents viscous damping, $\mu$ is the mass ratio between each metronome and the platform, and $\epsilon$ is a small parameter characterizing weak nonlinearity and weak coupling.

The function $g(x,\dot x)$ models the escapement mechanism, namely the driving force that compensates dissipation and sustains oscillations, and is given by
\begin{equation} 
g(x,\dot x)= \begin{cases} 
a x^3 - b x^5, & x\dot x > 0,\\ 0, & \text{otherwise}. \end{cases} 
\end{equation}

This simplified self-sustained oscillator model captures characteristic phenomena such as the coexistence of oscillatory and resting states and feedback-induced oscillation quenching while remaining analytically tractable.

For generating the artificial data, independent dynamical Gaussian noise with strength $\sigma$ is added to each oscillator.
The oscillator phase is extracted from the state variables by embedding the trajectory in the phase space $(x_i,\dot{x}_i)$ and defining the phase as the continuously unwrapped polar angle 
$\theta_i := \arg(x_i - i\dot{x}_i)$.

\begin{figure}[t]
  \centering
  \includegraphics[width=\linewidth]{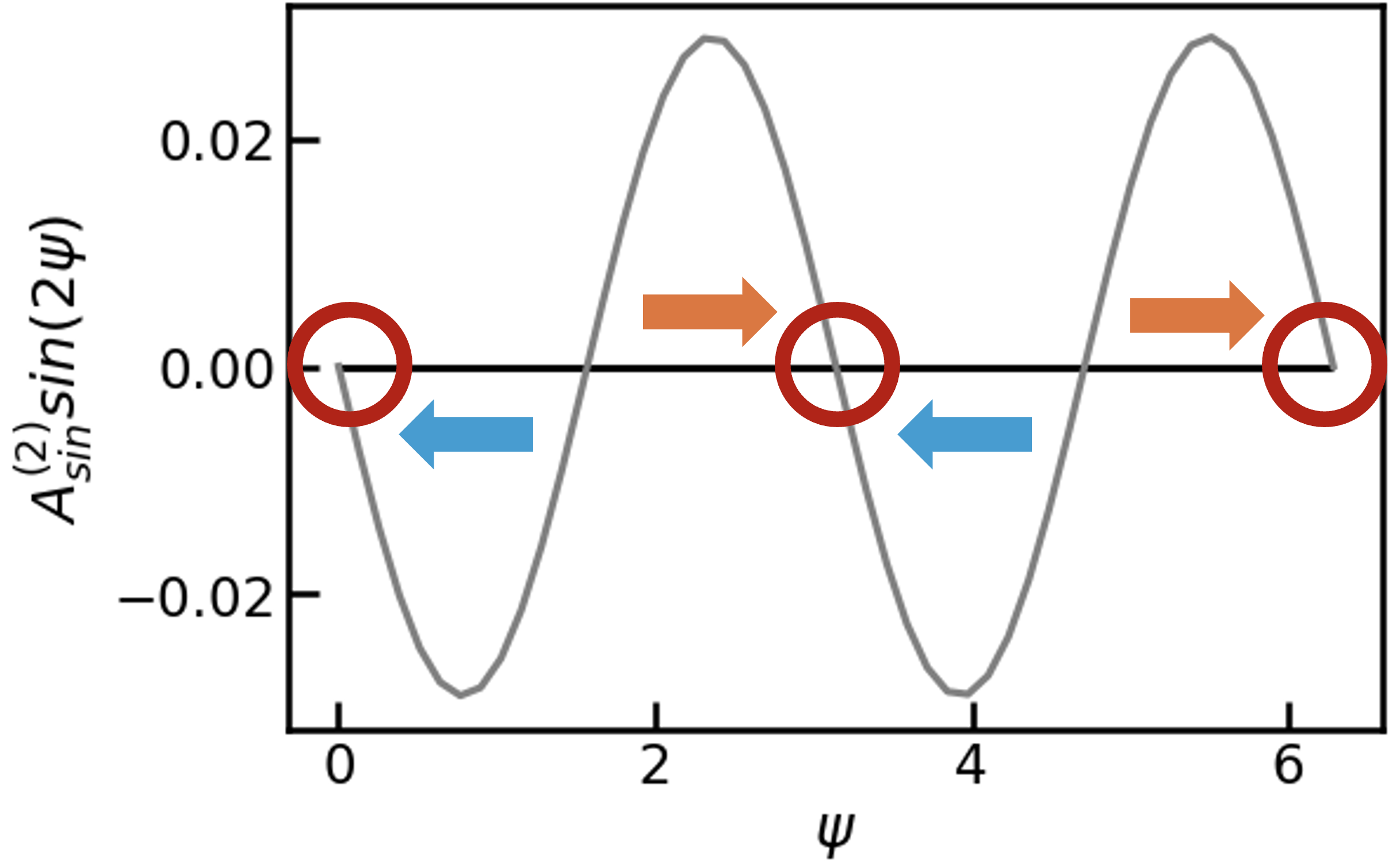}
  \caption{Plot of the inferred phase coupling term $A_{\sin}^{(2)} \sin(2\psi)$.
The coefficient $A_{\sin}^{(2)}$ is estimated from the data using the time window $[0,1000]$.
Red circles indicate the fixed points of the inferred dynamical system.}
  \label{fig:basin}
\end{figure}

We analyze the dynamics of the phase difference 
$\psi := \theta_1 - \theta_2$ 
using the following basis expansion:
\begin{eqnarray}
\frac{d\psi}{dt}
= \sum_{l=1}^{L} 
c^{(l)}_{\sin} A^{(l)}_{\sin}\sin(l\psi)
+
\sum_{l=1}^{L} 
c^{(l)}_{\cos} B^{(l)}_{\cos}\cos(l\psi).
\end{eqnarray}

For each basis function, we introduce binary indicator variables 
$c^{(l)}_{\sin}$ and $c^{(l)}_{\cos}$ that determine whether the corresponding 
basis component is included in the model.
This formulation is a special case of the general framework introduced in 
Eq.~(\ref{general_formulation}), where each basis function, coefficient, and indicator corresponds 
to $g_{i,m}^{(\gamma)}$, $\Theta_i^{(\gamma)}$, and $c_i^{(\gamma)}$, 
respectively.

Figure~\ref{fig:phase_diff} shows the phase difference $\psi$ extracted from the simulated 
metronome trajectory over the time interval $t \in [0,1000]$.
To investigate how the amount of data affects the inferred model, 
we analyze the trajectory using time windows of different lengths:
(1) $t \in [0,300]$, 
(2) $t \in [0,500]$, and 
(3) $t \in [0,1000]$.

Figures~\ref{fig:d_metro} show the posterior inclusion probabilities for each basis 
function obtained from the Bayesian analysis.
For the shortest time window (Fig.~\ref{fig:d_1_metro}), all basis functions exhibit 
low inclusion probabilities, indicating that the available data are 
insufficient to determine the effective phase dynamics.
When the time window is extended (Fig.~\ref{fig:d_2_metro}), the inclusion probability 
of the basis $\sin(2\psi)$ increases compared with the previous case.
Furthermore, when the full dataset is used (Fig.~\ref{fig:d_3_metro}), the inclusion 
probability of $\sin(2\psi)$ becomes even larger and clearly dominates 
the other basis functions.

The basis $\sin(2\psi)$ plays an essential role in representing bistability 
in the phase difference dynamics.
Figure~\ref{fig:basin} shows the inferred contribution of this component, 
$A_{\sin}^{(2)} \sin(2\psi)$, where the coefficient $A_{\sin}^{(2)}$ 
is estimated according to Eq.~(\ref{eq:coeff_esti}) within the Bayesian framework.
As shown in Fig.~\ref{fig:basin}, the function $A_{\sin}^{(2)} \sin(2\psi)$
exhibits two zero crossings within the interval $[0,2\pi]$,
across which the contribution of this term to $d\psi/dt$
changes sign from positive to negative.
These sign changes correspond to two stable configurations of the phase 
difference located near $\psi = 2m\pi$ and $\psi = (2m+1)\pi/2$ for 
integer $m$, which is consistent with the bistable synchronization states 
observed in the metronome system.

The results indicate that the posterior inclusion probability associated 
with the bistability-generating term increases as the amount of data grows.
Although the statistical model used for inference does not captures only a simplified abstraction of 
the true mechanical dynamics of the metronome system, the proposed Bayesian 
framework is nevertheless able to extract an effective phase model and 
identify the essential functional components with quantified uncertainty.

\section{Discussion and Conclusion}

In this study, we developed a Bayesian framework for identifying governing
equations of dynamical systems based on sparse modeling, together with
uncertainty quantification using Bayesian model averaging (BMA).
The proposed method enables the estimation of interaction structures in
dynamical systems with many possible couplings, as well as the selection of
relevant functional components from a large set of candidate basis functions,
while providing posterior inclusion probabilities that quantify the
credibility of each inferred component.

Through numerical experiments on oscillator networks, we demonstrated that
the proposed method can accurately recover interaction structures and
functional components with high credibility in asynchronous regimes.
In contrast, when the system exhibits phase-locked dynamics, the estimation
performance deteriorates due to the effective linear dependence among basis
functions. Importantly, however, the Bayesian formulation allows us to
quantify the resulting uncertainty through low posterior inclusion
probabilities. Furthermore, we showed that introducing moderate dynamical
noise breaks the phase locking, increases dynamical variability, and
significantly improves both estimation accuracy and credibility.

We also applied the proposed framework to a data-driven phase reduction of a
mechanical metronome system. In this case, the true governing equations of
the data-generating system do not belong to the assumed statistical model.
Nevertheless, the analysis successfully identified the functional component
responsible for bistability in the phase difference dynamics.
In particular, the posterior inclusion probability associated with the
$\sin(2\psi)$ component increases as the amount of data grows, indicating
that the essential nonlinear structure of the effective phase model can be
extracted from the data.

The ability to quantify the credibility of inferred components is crucial in
phenomenological modeling of dynamical systems. In many practical situations,
such as phase-locked dynamics, the governing functions may be inherently
difficult to identify from the available data. In addition, in settings such
as data-driven phase reduction, the statistical model used for inference may
not coincide with the true underlying dynamics that generate the data.
In such cases, point estimates alone are insufficient to determine whether
the inferred model is reliable. The uncertainty quantification provided by
Bayesian model averaging therefore plays an important role in assessing the
validity of phenomenological models constructed from data.

A limitation of the present approach is its computational cost.
In interaction structure inference, the number of candidate parameters grows
rapidly with the order of interactions: for pairwise interactions the number
scales as $O(N^2)$, while for three-body interactions it scales as $O(N^3)$.
Sampling such high-dimensional model spaces using parallel tempering
therefore becomes computationally demanding.

Developing methods to reduce the computational cost is an important direction
for future work. Possible approaches include approximate Bayesian inference
methods such as variational Bayes \cite{bishop2006pattern} or belief propagation \cite{mezard2009information}.
Another promising direction is to develop coarse-grained descriptions that
represent the collective dynamics of large oscillator networks using a
smaller number of effective variables \cite{izumida2013coarse}.
Advances along these lines would further extend the applicability of
Bayesian sparse modeling to large-scale nonlinear dynamical systems from
both statistical and physical perspectives.

\begin{acknowledgments}
We thank Ryota Kobayashi and Masaumi Shimizu for valuable comments.
This work is supported by JSPS KAKENHI (23KJ0723, 23H00486, JP23KJ0756).
We also thank the Kashiwa Campus Soccer Club for providing the opportunity for this collaboration.
\end{acknowledgments}

\appendix

\nocite{*}

\bibliography{papers.bib}

\end{document}